\documentclass[final,superscriptaddress,english,twocolumn,amssymb,nobibnotes,aps,longbibliography]{revtex4-1}
\usepackage{graphicx}
\usepackage{natbib}
\usepackage{amsmath}
\usepackage{amssymb}
\usepackage{appendix}
\usepackage[dvipsnames]{xcolor}{\huge }
\usepackage[section]{placeins}
\definecolor{darkblue}{rgb}{0,0,.65}
\definecolor{darkgreen}{rgb}{0.28,0.41,0.19}
\usepackage{pifont}
\usepackage{wrapfig}{\LARGE }
\usepackage{nicefrac}
\usepackage[%
    pdfauthor={Robin Schaefer},%
  pdfstartview=FitH,%
  breaklinks=true,%
  bookmarks=true,%
  colorlinks=true,%
  anchorcolor=black,%
  citecolor=blue,
  filecolor=black,%
  menucolor=black,%
  urlcolor=darkblue,%
  linkcolor=blue,%
 ]{hyperref}
\usepackage[all]{hypcap} %
\newcommand{\bra}[1]{\langle\,#1\,|}
\newcommand{\ket}[1]{|#1\rangle}

\newcommand{\braket}[2]{\langle\,#1\, | \, #2\,\rangle}
\newcommand{\braOket}[3]{\langle\,#1\, | \, #2 \, | \, #3\,\rangle}

\newcommand{\cf}{\textit{cf.} }

\newcommand{\eg}{\textit{e.g.} }

\graphicspath{{images/}}
\begin{document}

\title{Symmetry protected exceptional points of interacting fermions}

\author{Robin Sch\"afer}\email{schaefer@pks.mpg.de}
\affiliation{Max Planck Institute for the Physics of Complex Systems, Noethnitzer Str. 38, 01187 Dresden, Germany}
\author{Jan C. Budich}\email{jan.budich@tu-dresden.de}
\affiliation{Institute of Theoretical Physics, Technische Universit\"{a}t Dresden and W\"{u}rzburg-Dresden Cluster of Excellence ct.qmat, 01062 Dresden, Germany}
\author{David J. Luitz}\email{david.luitz@uni-bonn.de}
\affiliation{Institute of Physics, University of Bonn, Nussallee 12, 53115 Bonn, Germany}
\affiliation{Max Planck Institute for the Physics of Complex Systems, Noethnitzer Str. 38, 01187 Dresden, Germany}

\date{\today}

\begin{abstract}
    Non-hermitian quantum systems can exhibit spectral degeneracies known as \emph{exceptional points}, where two or more eigenvectors coalesce, leading to a non-diagonalizable Jordan block. 
    It is known that symmetries can enhance the abundance of exceptional points in non-interacting systems. 
    Here, we investigate the fate of such symmetry protected exceptional points in the presence of a symmetry preserving interaction between fermions and find that, (i) exceptional points are stable in the presence of the interaction. Their propagation through the parameter space leads to the formation of characteristic exceptional ``fans''. 
	In addition, (ii) we identify a new source for exceptional points which are only present due to the interaction. These points emerge from diagonalizable degeneracies in the non-interacting case.
	Beyond their creation and stability, (iii) we also find that exceptional points can annihilate each other if they meet in parameter space with compatible many-body states forming a \emph{third} order exceptional point at the endpoint. These phenomena are well captured by an ``exceptional perturbation theory'' starting from a non-interacting Hamiltonian.
\end{abstract}

\maketitle

Dissipative phenomena in physics have been effectively described by non-hermitian Hamiltonians \cite{carmichael_quantum_1993,breuer_theory_2007,rotter_non-hermitian_2009,bender_making_2007,gardiner_quantum_2004,brody_biorthogonal_2013,daley_quantum_2014,ashida_non-hermitian_2020} in a wide range of settings, including photonic systems \cite{brandstetter_reversing_2014,zhen_spawning_2015,doppler_dynamically_2016,cerjan_experimental_2019,lee_observation_2009,cao_dielectric_2015,hahn_observation_2016,peng_chiral_2016,chen_exceptional_2017,choi_quasieigenstate_2010,bandres_topological_2018,xu_topological_2016,zeuner_observation_2015} and correlated electron systems \cite{lehmann_dynamically_2021,michishita_relationship_2020,yoshida_exceptional_2020,kimura_chiral-symmetry_2019,yoshida_real-space_2021,kozii_non-hermitian_2017,shen_quantum_2018,zyuzin_flat_2018,papaj_nodal_2019,yoshida_non-hermitian_2018,mitscherling_non-hermitian_2021,okuma_non-hermitian_2021,yoshida_exceptional_2020,nagai_dmft_2020,yoshida_real-space_2021,ferry_open_2013,hyart_non-hermitian_2022,zhang_transport_2018,mortemousque_enhanced_2021}.
Several recent studies have investigated properties of \emph{interacting} non-hermitian Hamiltonians \cite{fukui_breakdown_1998,mu_emergent_2020,castro-alvaredo_spin_2009,pan_interacting_2019,xu_topological_2020,bebiano_quantum_2020,aquino_exceptional_2020,aquino_probing_2021,yamamoto_theory_2019,rausch_exceptional_2021,crippa_fourth-order_2021}, in particular critical fluctuations \cite{hanai_critical_2020}, coupled quantum dots \cite{ferry_open_2013,hyart_non-hermitian_2022,zhang_transport_2018,mortemousque_enhanced_2021}, the topology of many-body spectra \cite{luitz_exceptional_2019} or non-hermitian many-body localization \cite{medvedyeva_influence_2016,levi_robustness_2016,hamazaki_non-hermitian_2019,zhai_many-body_2020} but also the appearance of non-hermitian physics in phenomena such as magnon decay \cite{mcclarty_non-hermitian_2019}.

\begin{figure}[t!]
	\centering
	\includegraphics[width=\columnwidth]{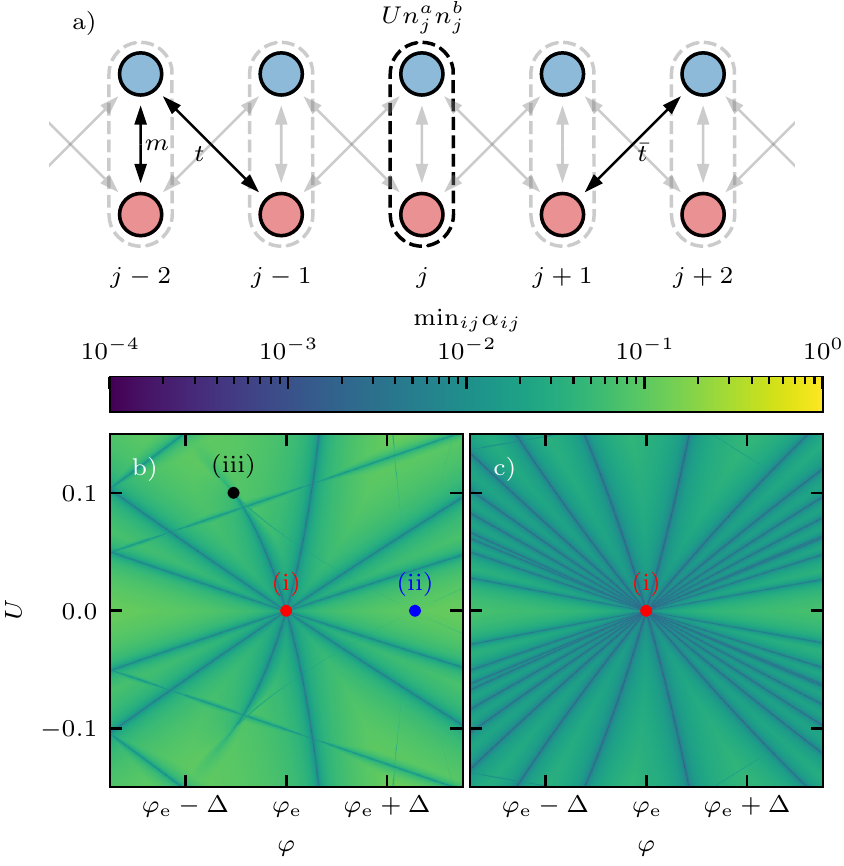}\label{fig:heatmap}
	\caption{a) Non-hermitian Hamiltonian with hopping $t=e^{i3/4\pi}/\sqrt{2}$ and interaction $U\in\mathbb{R}$. The hopping $j\rightarrow j+1$ picks up a phase defined by the twist angle $\varphi\in[0,2\pi)$. b) and c) Propagation of EPs of two interacting fermions (dark lines) as a function of $\varphi$ ($\Delta = 0.02$) and $U$ for sizes $L=6$ (b)) and $L=18$ (c)). EPs forming the ``fan'' feature (i) emanate from the same symmetry protected EP for $m=0.7$ at $\varphi_{\text{e}}$ (cf. Fig. \ref{fig:1BZ}). EPs can emerge from diagonalizable degeneracies (ii), and annihilate each other (iii). The color scale indicates the minimal angle  $\alpha_{ij}=\arccos( |\braket{\Psi_i^R}{\Psi_j^R}|)$ between two right eigenvectors.\vspace{0.75cm}}
\end{figure}

There, exceptional points (EPs) \cite{heiss_exceptional_2004,uzdin_observability_2011,heiss_physics_2012,okugawa_topological_2019,zhou_exceptional_2019,huang_exceptional_2020,holler_non-hermitian_2020,budich_symmetry-protected_2019,yoshida_symmetry-protected_2019,sayyad_realizing_2022}, i.e. spectral degeneracies at which also two (second order EPs) or more (higher order EPs) eigenvectors coalesce so as to render the Hamiltonian non-diagonalizable, represent the generic counterpart of level crossings familiar from hermitian systems. EPs are more abundant than diagonalizable degeneracies, and thus become the rule rather than the exception as soon as dissipative sources of non-hermiticity enter the stage. Specifically, two real parameters need to be tuned to find a second order EP (co-dimension two), while three real parameters in hermitian systems and even six in non-hermitian systems are required to yield a diagonalizable degeneracy. Notably, symmetries such as chiral and PT symmetry further reduce the co-dimension of EPs by a factor of two, rendering second order symmetry protected EPs topologically stable in one-dimensional systems \cite{budich_symmetry-protected_2019,yoshida_symmetry-protected_2019,delplace_symmetry-protected_2021}.

In this work, we analyze the fate of PT symmetry protected EPs in non-hermitian Bloch bands in the presence of both repulsive and attractive two-body interactions with strength $U$, \cf Fig.~\ref{fig:heatmap}a). To this end, the single particle lattice momentum, acting as the tuning parameter for EPs in the non-interacting limit is generalized to a flux-variable $\varphi$ in the framework of twisted boundary conditions. Twisted boundary conditions appear naturally in one dimensional systems with periodic boundaries (\eg a closed ring) in the presence of a magnetic field. In the resulting $\varphi$-$U$ parameter plane, beams of EPs are emanating from their non-interacting origin (see point (i) in Fig.~\ref{fig:heatmap}b)), marking their stability under symmetry-preserving correlations. Beyond this mere robustness, we exemplify and explain theoretically how new EPs are induced by interactions from accidental diagonalizable degeneracies (see point (ii) in Fig.~\ref{fig:heatmap}b)). Finally, we find that pairs of EPs in the same total momentum sector can undergo a pairwise annihilation process (see point (iii) in Fig.~\ref{fig:heatmap}b)). Our numerical results are well captured by a non-standard perturbative expansion around the degeneracies \cite{kato_perturbation_1995,marie_perturbation_2021,znojil_perturbation_2020,bender_large-order_1999,sternheim_non-hermitian_1972,buth_non-hermitian_2004,castro-alvaredo_spin_2009,sun_biorthogonal_2021}.

We expect these results to be of relevance for a broad class of physical scenarios, where dissipative processes such as single particle gain or loss give rise to an effective non-hermitian band structure, while many-body scattering processes are well described by hermitian density-density interactions.

We start by introducing the model in Sec. \ref{sec:model} and discuss our results in a \emph{short} summary in Sec. \ref{sec:results} which is organized in three subsections. A detailed mathematical derivation and the perturbative approach can be found in the appendix. Sec. \ref{sec:con} summarizes our work and points towards further directions in the field.

\section{Model}\label{sec:model}
As illustrated in Fig. \ref{fig:heatmap}a), we investigate a one dimensional fermionic two-band model with sub lattices $a$ and $b$ and a complex hopping amplitude. In the non-interacting limit, $U=0$, we can derive the non-hermitian Bloch Hamiltonian
\begin{equation}
	H_0 = \sum_{k=0}^{L-1} \begin{pmatrix} a^\dagger_k & b_k^\dagger \end{pmatrix} 
	\begin{pmatrix} 0 & m_k \\ p_k & 0 \end{pmatrix} 
	\begin{pmatrix} a_k \\ b_k \end{pmatrix},  \label{eq:H0_model}\\
\end{equation}
    where $m_k, p_k\in\mathbb{R}$ are defined with $m\in \mathbb{R}$ by 
    \begin{equation*}
        \begin{split}
        m_k &= m - \cos\left(\frac{2\pi}{L} k + \frac{\varphi}{L}\right) - \sin\left(\frac{2\pi}{L} k + \frac{\varphi}{L}\right),\\
        p_k &= m - \cos\left(\frac{2\pi}{L} k + \frac{\varphi}{L}\right) + \sin\left(\frac{2\pi}{L} k + \frac{\varphi}{L}\right).
\end{split}
\end{equation*}

\begin{figure}
	\centering
    \includegraphics[]{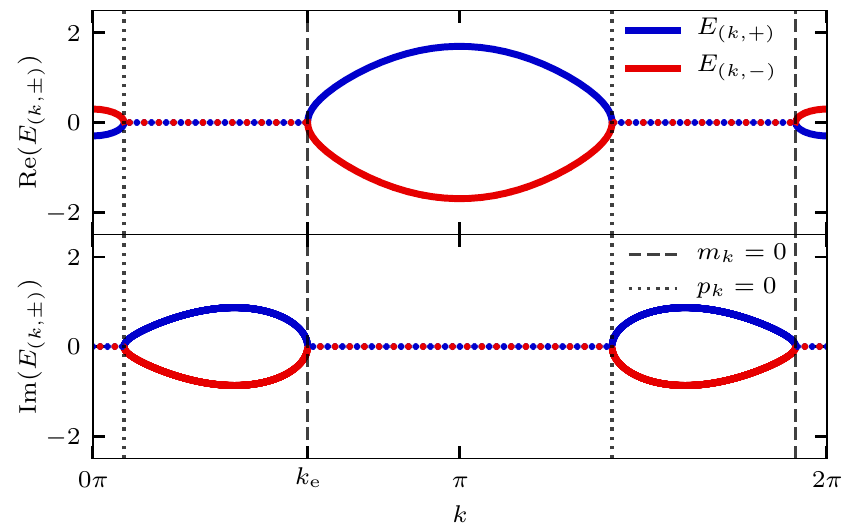}\label{fig:1BZ}
	\caption{Complex eigenenergies $E_{(k,\pm)}$, Eq. \eqref{eq:eigenvalues_main}, of the single particle Bloch Hamiltonian in Eq. \eqref{eq:H0_model} for $m=0.7$. For simplicity, we show here continuous momenta of an infinite chain, so that we do not need to rely on the twist angle $\varphi$ here. 
	$k_{\text{e}}= 2\arctan((10+\sqrt{151})/17)$ indicates the single particle EP shown in Fig. \ref{fig:heatmap}.  }
\end{figure}

Since a finite system only has a discrete set of $k$ points, we use twisted boundary conditions with twist angle $\varphi\in[0,2\pi)$, which allows us to continuously tune the momentum grid of $k$ points, and defines a counterpart of single particle momentum that generalizes to correlated many-body systems.
This model is time reversal and lattice inversion symmetric ($k\to-k$ and $a\leftrightarrow b$), and preserves the particle number $n=n^a+n^b$, thus ensuring the existence of EPs in $k$ space by symmetry \cite{budich_symmetry-protected_2019}. The symmetry is given by $H^* = IqHq^\top I^\top$, where $q$ inverts both orbitals and $I$ inverts all sites in real space. The Bloch matrix in Eq. \eqref{eq:H0_model} becomes non-diagonalizable if either $m_{k_{\text{e}}}=0$ or $p_{k_{\text{e}}}=0$. 
This happens at four points in the first Brillouin zone
\begin{equation}
    \frac{2\pi k_{\text{e}} +\varphi_{\text{e}}}{L} = 2 \arctan\left( \frac{ \pm \sqrt{2-m^2} \pm 1 }{m+1} \right) \mod 2\pi,
    \label{eq:EP0_cond}
\end{equation}
yielding four solutions (the two sign choices are independent) for EPs tunable by $\varphi_\text{e}$ at a given momentum $k_\text{e}$.
For each (shifted) momentum $(2\pi k + \varphi)/L$, $H_0$ has two single particle eigenvalues 
\begin{equation}
	E_{(k,\pm)} = \pm \sqrt{p_k} \sqrt{m_k} \in \mathbb{C}
    \label{eq:eigenvalues_main}
\end{equation}
shown in Fig. \ref{fig:1BZ}. 
Since $m_k$ and $p_k$ switch signs at their zeros, the eigenvalues of $H_0$ are either real or imaginary and the switch between these two cases occurs at the EPs.

We can represent the corresponding right eigenstates $\ket{E_{(k,\pm)}^R}$ and left eigenstates $\bra{E_{(k,\pm)}^L}$ in the single particle basis in momentum space spanned by $2L$ states:
\begin{align}
	\ket{ E_{(k,\pm)}^R} &= \left(\pm \sqrt{m_k}  a_k^\dagger + \sqrt{p_k} b_k^\dagger \right)\ket{0}/\sqrt{ 2E_{(k,+)}}\\
	\langle  E_{(k,\pm)}^L \vert &= \bra{0} \left(\pm \sqrt{p_k} a_k + \sqrt{m_k} b_k \right)/\sqrt{ 2E_{(k,+)}}
\end{align}

Away from EPs, the  left and right eigenvectors satisfy the orthogonality relation $\braket{E_{(k,\xi_k)}^L}{E_{(q,\xi_q)}^R} = \delta_{kq}\delta_{\xi_k\xi_q}$, $\xi_k,\xi_q=\pm$. The model is defective if $E_{(k_{\text{e}},\pm)}=0$ where both eigenvectors align. We drop the specific distinction between left and right eigenvectors and absorb it in the \textit{bra}-\textit{ket} notation.

In this paper, we are interested in the fate of the symmetry protected EPs defined by Eq. \eqref{eq:EP0_cond} in the presence of both attractive ($U<0$) and repulsive ($U>0$) \emph{interactions}. We consider a simple density-density interaction, which preserves the symmetries of the system and is \emph{hermitian}:
\begin{equation}
    H_\text{int} = \sum_{j=0}^{L-1} n_j^a n_j^b=\frac{1}{L} \sum_{k,k',q}^{L-1} a_k^\dagger a_{k+q} b^\dagger_{k'} b_{k'-q}.
    \label{eq:interaction}
\end{equation}

We consider the Hamiltonian $H = H_0 + U H_\text{int}$ in the simplest non-trivial case of \emph{two interacting fermions} with a Hilbert space dimension $D=L(2L-1)$.

\section{Results}\label{sec:results}
We start by a numerical characterization of EPs as a function of the twist angle $\varphi$ and interaction strength $U$ for two fermions in a finite system of length $L$. For each parameter set $(\varphi,U)$, we calculate all right eigenstates of the two-particle Hamiltonian numerically. If we are close to an EP, two eigenvectors will align, enclosing a very small angle. We have found that a robust quantifier for the identification of EPs is therefore to consider the \emph{smallest angle} $\min_{ij} \alpha_{ij} = \min_{ij} \arccos( |\braket{\Psi_i^R}{\Psi_j^R}|)$ enclosed by any pair of right eigenvectors $\ket{\Psi_i^R}$ and $\ket{\Psi_j^R}$.

Fig. \ref{fig:heatmap} shows $\min_{ij} \alpha_{ij}$ in the parameter plane, exhibiting sharp lines of very small angles (dark lines), which we identify as EPs. For $U=0$ we recover the non-interacting model with EPs at $(2\pi k_{\text{e}} + \varphi_{\text{e}})/{L}$, where in a finite chain the EP is located in the momentum sector $k_{\text{e}}$ and realized at twist angle $\varphi_{\text{e}}$. At an EP, both aligned single particle states can be combined with any non-exceptional state $\ket{E_{(q,\pm)}}$ forming two identical two-particle wavefunctions yielding $2L-2$ two dimensional Jordan blocks in the non-interacting case.

Fig. \ref{fig:heatmap} displays a rich phenomenology of EPs: (i) The EPs from the non-interacting case extend into \emph{exceptional lines} in the form of a ``fan'' for finite interaction strength; (ii) At special points, where the non-interacting model exhibits a diagonalizable degeneracy, EPs can emerge in the presence of a \emph{hermitian} interaction ($U\neq0$), creating very sharp ``lines''; (iii) When two lines of EPs meet in the parameter space, they can extinguish and form an endpoint in the case of a hermitian interaction, $U\in\mathbb{R}$.

In the following we will discuss these three phenomena in detail using non-hermitian degenerate perturbation theory for a small interaction strength $U$. 
Starting from the non-interacting limit, we identify all eigenstates which have degenerate eigenenergies for $U=0$ and create the effective Hamiltonian in the space spanned by the corresponding \textit{generalized} right and left eigenvectors. Since the total Hamiltonian is translational invariant, the effective Hamiltonian can be reduced to blocks with fixed total momentum.
We then derive constrains for $U(\varphi)$ such that the effective Hamiltonian is non-diagonalizable.

\subsection{(i) Robustness of EPs}  

Fig. \ref{fig:heatmap} illustrates that the EP stemming from a non-diagonalizable Bloch Hamiltonian for momentum $k_\text{e}$ with twist angle $\varphi_\text{e}$ at $U=0$ is robust if the interaction $U$ is turned on.
We will therefore focus on eigenstates corresponding to the defective blocks of the two-particle Hamiltonian in the non-interacting limit.
The eigenenergies of the two-particle Hamiltonian are sums of one particle eigenenergies. The energy at a single particle EP is zero, $E_{(k_\text{e},\pm)}=0$, and both eigenvectors coalesce to $\ket{a_{k_\text{e}}}= a_{k_\text{e}}^\dagger \ket{0}$ (for $p_{k_\text{e}}=0$) and $\ket{b_{k_\text{e}}}=b_{k_\text{e}}^\dagger \ket{0}$ (for $m_{k_\text{e}}=0$). 
The coalescing eigenvectors can be combined with any non-exceptional state, $\ket{E_{(q,\pm)}}$, to form two identical two-body wavefunctions exhibiting the same eigenvalue $E_{(q,\pm)}$ (since $E_{(k_\text{e},\pm)} =0$).

Since at the EP the Hamiltonian is defective and the only eigenvector does not span the full space corresponding to the two fold degenerate eigenvalue, we need to represent the effective Hamiltonian in the space spanned by the two generalized eigenvectors with eigenvalue $E_{(q,\pm)}$, $\ket{a_{k_\text{e}}; E_{(q,\pm)}}$ and  $\ket{b_{k_\text{e}}; E_{(q,\pm)}}$. The generalized eigenvectors span the space of the Jordan block and satisfy $(H-E_{(q,\pm)})^2 \ket{a_{k_\text{e}}; E_{(q,\pm)}} = 0$ (and respectively for $\ket{b_{k_\text{e}};E_{(q,\pm)}}$).
Additional accidental degeneracies are practically impossible in the same momentum sector for a system of finite size.

Calculating matrix elements between left and right generalized eigenvectors we obtain the effective Hamiltonian
\begin{equation}
	H^{\text{(i)}} = \begin{pmatrix} E_{(q,\pm)}& m_{k_{\text{e}}} \\ p_{k_{\text{e}}} & E_{(q,\pm)}\end{pmatrix} +  \frac{U}{2L}\begin{pmatrix} 1 &  \mp\sqrt{\nicefrac[\texttt]{$m_q$}{$p_q$}}  \\ \mp \sqrt{\nicefrac[\texttt]{$p_q$}{$m_q$}} & 1 \end{pmatrix}.
\end{equation}
We are now interested if and at which finite interaction strength $U$ and twist angle $\varphi$ this matrix remains defective. Since the diagonal entries of $H^\text{(i)}$ are equal this happens if and only if $H^{\text{(i)}}_{01}=0$ or $H^{\text{(i)}}_{10}=0$ which yields the conditions
\begin{align}
	U_{(q,\pm)}^m = \pm{2L}{m_{k_{\text{e}}}}\frac{\sqrt{p_q}}{\sqrt{m_q}}\text{ and }U_{(q,\pm)}^p = \pm{2L}{p_{k_{\text{e}}}}\frac{\sqrt{m_q}}{\sqrt{p_q}}\label{eq:eps_stability}
\end{align}
If the non-interacting Hamiltonian has an EP generated from $\ket{E_{(q,\pm)}}$ for $m_{k_{\text{e}}}=0$ ($p_{k_{\text{e}}}=0$) it propagates through the parameter space according to $U_{(q,\pm)}^m$ ($U_{(q,\pm)}^p$). 
The EPs are only preserved for a hermitian interaction if $U_{(q,\pm)}^m$ ($U_{(q,\pm)}^p$) is real which is equivalent to $E_{(q,\pm)}$ being real. For imaginary $E_{(q,\pm)}$, EPs instead survive only in the presence of an anti-hermitian interaction, $U\in i\mathbb{R}$ (\cf appendix). This explains why the number of exceptional lines visible in Fig. \ref{fig:heatmap} is not $2L-2$ but only roughly $\approx L$. Our analytical prediction from Eq. \eqref{eq:eps_stability} is shown in comparison with the numerical result in Fig. \ref{fig:EP_prediction}a) with excellent agreement. Additionally, our perturbative treatment allows us to determine the exceptional eigenvector which remains in the state $\ket{a_{k_\text{e}};E_{(q,\pm)}}$ or $\ket{b_{k_\text{e}};E_{(q,\pm)}}$ for a finite interaction strength.

\subsection{(ii) Emergence of EPs} We identify a new source of EPs which is only present in the case of an interacting many-body system. It has been shown \cite{luitz_exceptional_2019} that EPs can emerge from a non-hermitian interaction. Here, we show that EPs can also emerge from a diagonalizable degeneracy of our non-interacting model ($U=0$) in the presence of a hermitian interaction. 

A common source for degeneracies in the case of two fermions is induced by degeneracies in the single particle spectrum with different momenta ($k\neq q$), $E_{(k,+)}=E_{(q,-\xi)}$, $\xi=\pm$, at $\varphi_\text{d}$ for $U=0$. Since each eigenvalue comes with either sign, this produces pairs of two-particle states ($\ket{\Psi_+}=\ket{E_{(k,+)};E_{(q,\xi)}}$ and $\ket{\Psi_-}=\ket{E_{(k,-)};E_{(q,-\xi)}}$) in the total momentum sector $k+q$ with eigenvalues 
\begin{align}
	\delta=E_{(k,+)} + E_{(q,\xi)}\text{ and } -\delta = E_{(k,-)} + E_{(q,-\xi)}
\end{align}
which coalesce to \textit{zero} at $\varphi_\text{d}$. A full perturbative description needs to include all states exhibiting the same energy, here zero, at $\varphi_\text{d}$. Depending on the length $L$ and total momentum $k+q$, additional states $\ket{\Phi_p}=\ket{E_{(p,+)};E_{(p,-)}} = a^\dagger_p b^\dagger_p \ket{0}$ with eigenvalue zero exist, which need to be included in the perturbative subspace. Hence, we extend the effective Hamiltonian, $H^{\text{(ii)}}$, with zero (even $L$, odd $k+q$), one (odd $L$) or two (even $L$, even $k+q$) additional states. Similarly to the case of EPs which are inherited from the single particle spectrum we can derive conditions for a non-diagonalizable effective Hamiltonian depending on the size of the subspace:

\begin{align}
	U^\pm_{2\times 2} &= \pm i\frac{L\delta}{a^2}\label{eq:EP_2x2}\\
	U^{\pm,\pm}_{3\times 3} &= \pm L\delta\sqrt{\xi\frac{a^4 -\xi 10 a^2 - 2 \pm a\sqrt{(a^2+4\xi )^3}}{2(2a^2-\xi)^3}}\label{eq:EP_3x3}\\
	U^{\pm,\pm}_{4\times 4} &= \pm L\delta\sqrt{\xi\frac{a^4 -\xi 20 a^2 - 8 \pm a\sqrt{(a^2+8\xi )^3}}{32(a^2-\xi)^3}}\label{eq:EP_4x4}\\
	\text{with } a &=\frac{\sqrt{m_{k}p_{q}}-\xi \sqrt{p_{k}m_{q}} }{2\sqrt{E_{(k,+)}}\sqrt{E_{(q,+)}}}
\end{align}

\begin{figure}
	\centering
	\includegraphics[width=\columnwidth]{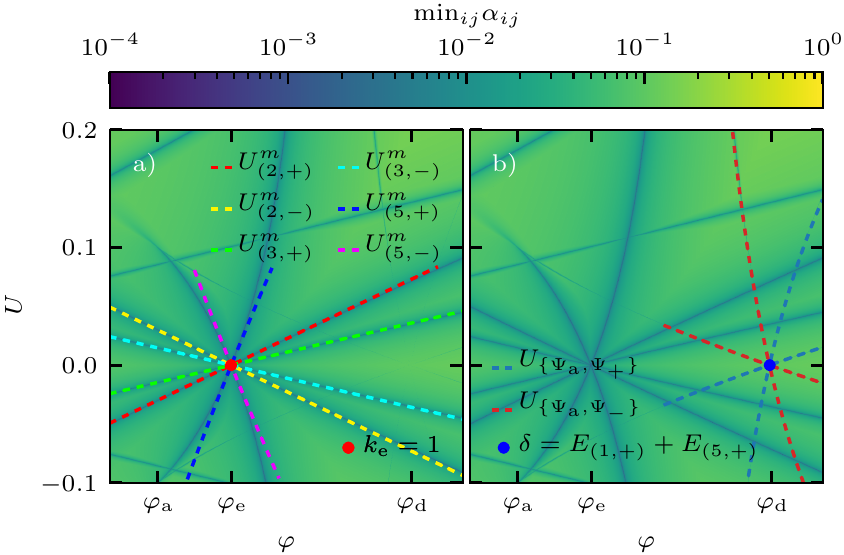}\label{fig:EP_prediction}
	\caption{Comparison of our analytic predictions (dashed lines) of EPs to the numerical simulation for $L=6$ and $m=0.7$. We identify three different twist angles where (i) EPs are inherited from the single particle spectrum at $\varphi_{\text{e}}$, (ii) emerge from a diagonalizable degeneracy at $\varphi_\text{d}$ and (iii) annihilate each other at $\varphi_\text{a}$. Panel a) and b) show the analytical predictions from Eq. \eqref{eq:eps_stability} and \eqref{eq:EP_4x4} and identify the states forming the exceptional lines.}
\end{figure}

Again, we find an excellent agreement of our prediction, here Eq. \eqref{eq:EP_4x4}, with the numerical simulation in Fig. \ref{fig:EP_prediction}b). The derived estimates can be used to evaluate the stability of EPs for a finite hermitian interaction.
Besides the prediction of a defective Hamiltonian in the parameter space, we are able to assign states forming the emergent EPs. The color code in the Fig. \ref{fig:EP_prediction}b) of the predicted paths refers to states $\{\Psi_\alpha,\Psi_\beta\}$ forming the EPs that can be adiabatically connected to the non-interacting limit. While $\ket{\Psi_\text{a}}$ refers to the additional state (either pure or as a superposition of two), $\ket{\Psi_+}$ and $\ket{\Psi_-}$ refer to $\ket{E_{(k,+)};E_{(q,\xi)}}$ and $\ket{E_{(k,-)};E_{(q,-\xi)}}$, respectively.
A detailed derivation and the effective Hamiltonians are given in the supplementary material.


\subsection{(iii) Annihilation} A careful inspection of the evolution of EPs at finite interaction strength reveals that certain pairs of EPs are annihilated if two exceptional lines meet in the parameter space (indicated by (iii) in Fig. \ref{fig:heatmap}). A first indicator to identify these pairs is the conserved total momentum which protects EPs by symmetry if they are located in different momentum sectors.
However, a second mechanism allows some lines of EPs in the same momentum sector to cross.

While both perturbative expansions, (i) and (ii), give a precise estimate around the non-interacting limit, they fail to resolve the annihilation process. Especially, EPs inherited from (i) which are later annihilated deviate from their analytic prediction suggesting that the perturbative subspace is insufficient. A complete description capturing all three phenomena has to include all states forming the EPs emerging from (i) and (ii) which can be extracted using the effective Hamiltonians $H^{\text{(i)}}$ and $H^{\text{(ii)}}$. We observe that exceptional lines form an endpoint if they are composed of the same single particle state $\ket{E_{(q,\xi)}}$, $\xi=\pm$ ($\ket{E_{(5,-)}}$ in Fig. \ref{fig:EP_prediction}). The EP inherited from (i) is generated by $\ket{a_{k_\text{e}};E_{(q,\xi)}}$ and $\ket{b_{k_\text{e}};E_{(q,\xi)}}$ and the EP emerging from (ii) is formed by $\ket{\Psi_\text{a}}$ and $\ket{\Psi_\pm} = \ket{E_{(k_\text{e},\pm)};E_{(q,\pm \xi)}}$. However, $\ket{\Psi_\pm}$ is a linear combination of the two states included in (i) and the full perturbative description can be reduced to a \emph{three} dimensional subspace. Hence, extending the effective description from (i) by the additional state $\ket{\Psi_\text{a}}$ from (ii) is sufficient to capture all three phenomena: the heredity of the EP (i), the emergence from a diagonalizable degeneracy (ii) and their annihilation at (iii) (\cf appendix). At the endpoint, both EPs of order \emph{two} coalesce and form a \emph{third} order EP such that the full effective Hamiltonian, $H^{\text{(iii)}}$, transforms into a Jordan block of size three.



Finally, even though the lines of EPs end in the case of a hermitian interaction ($U\in\mathbb{R}$) at the annihilation point, we show in the supplementary materials that they survive for a non-hermitian interaction ($U\in\mathbb{C}$).

\section{Conclusion}\label{sec:con}
We have shown that (i) symmetry protected EPs of a non-hermitian single particle Hamiltonian can persist in the presence of hermitian interactions between two fermions. Their precise location in the parameter space depends on the momenta of the involved particles. Furthermore, (ii) we identified a second source of EPs emerging from diagonalizable degeneracies in the non-interacting limit.
Besides the creation and stability of EPs, we observe that exceptional lines can annihilate each other, forming an endpoint (in the case of a hermitian interaction) if the involved many-body states are compatible. 
This phenomenology is captured with very high precision by non-hermitian perturbation theory, which predicts the location of EPs in the parameter space of two fermions.
Additionally, the perturbative treatment evaluates not only the stability of EPs in the case of a hermitian interaction but also suggests that EPs are restricted to their perturbative subspace. 
We have focused on the simplest case of two fermions here, but our findings can be generalized to the many-fermion limit as shown in the appendix \ref{app:three_fermions}.
Experiments suffer from disorder which break the translational invariance. 
Therefore, we evaluated the stability of EPs in the presence of disordered hopping amplitudes and found that the EPs still exist, but their behavior is more complex, \cf appendix \ref{app:disorder}.


\begin{acknowledgments}
    We are grateful to Francesco Piazza for valuable discussions and collaborations on related topics. This work was financially supported by the Deutsche Forschungsgemeinschaft through SFB 1143 (project-id 247310070), the cluster of excellence ct.qmat (EXC 2147, project-id 390858490) and the cluster of excellence ML4Q (EXC 2004, project-id 390534769). DJL acknowledges support from the QuantERA II Programme that has received funding from the European Union's Horizon 2020 research innovation programme (GA 101017733), and from the Deutsche Forschungsgemeinschaft through the project DQUANT (project-id 499347025).
\end{acknowledgments}

\bibliography{exceptional_points}
\appendix
\clearpage

\section{Model}
In the following, we are interested how the defective structure of the non-interacting Hamitonian influences a system with two fermions. We can extend the two-band model to two non-interacting fermions which is decomposed into $4\times 4$ blocks referring to two momenta $k\neq q$. These blocks inherit the defective structure occurring in the Bloch Hamiltonian for $k_{\text{e}}$ and $\varphi_{\text{e}}$. Each $4\times 4$ block containing $k_{\text{e}}$ becomes non-diagonalizable for $m_{k_{\text{e}}}=0$ or $p_{k_{\text{e}}}=0$ and can be transformed into two $2\times 2$ Jordan blocks with eigenvalue $E_{(q,\pm)}$. Hence, the single particle EP induces $2(L-1)$ Jordan blocks in the case of two fermions.
\begin{align}
	H_0 &= \sum_{k=0}^{L-1} \sum_{q=0}^{k-1} \vec{\Psi}_{k,q}^\dagger \begin{pmatrix} 0 & 0 & m_q & m_k \\ 0 & 0 & p_k & p_q \\ m_q & p_k & 0 & 0 \\ m_k & p_q & 0 & 0\end{pmatrix} \vec{\Psi}_{k,q} \label{eq:H0_model_2_fermions}\\
	&+ 0 \sum_{k=0}^{L-1} a_k^\dagger b_k^\dagger a_k b_k\nonumber
\end{align}
with $ \vec{\Psi}^\dagger_{k,q} =  \begin{pmatrix} a_k^\dagger a_q^\dagger & b_k^\dagger b_q^\dagger & a_k^\dagger b_q^\dagger & b_k^\dagger a_q^\dagger \end{pmatrix}$. Note that the Hamiltonian naturally exhibits trivial eigenstates with zero energy created by a single momentum. One central question of our work is the stability and behavior of the two-particle EPs in the $4\times 4$ block which are inherited from the defective Bloch Hamiltonian in the case of interacting fermions.


\section{Effective Hamiltonian}\label{app:states}
\begin{figure}
	\centering
	\includegraphics[width=\columnwidth]{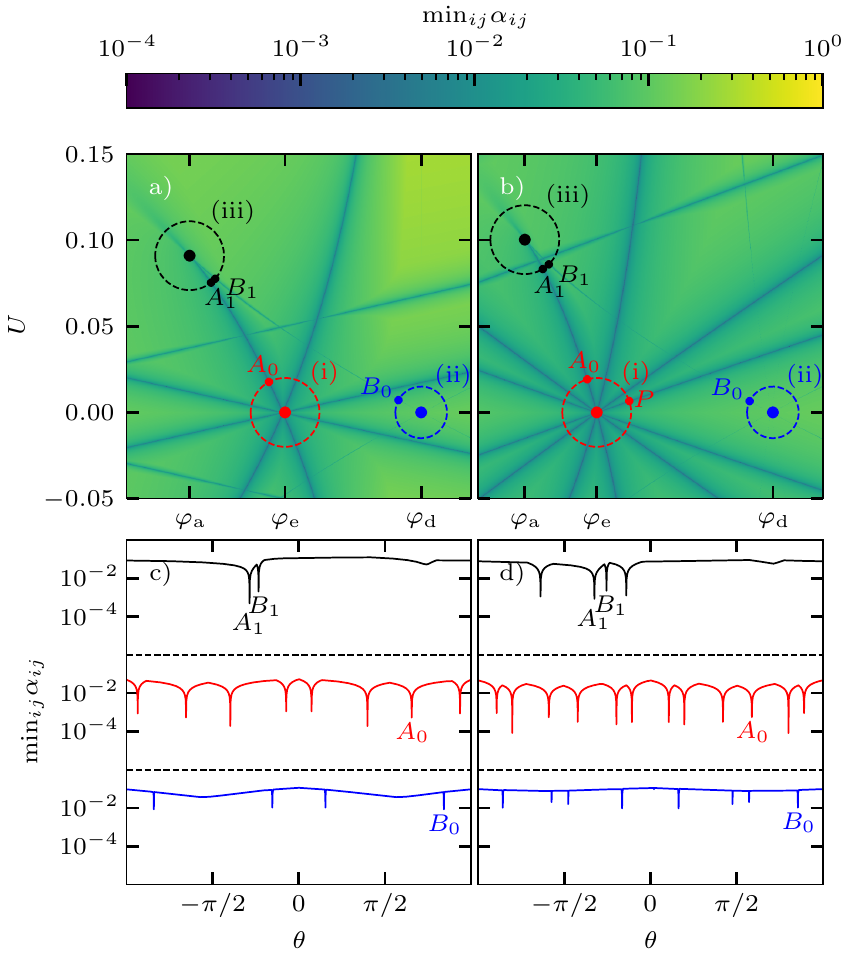}\label{fig:circles_app}
    \caption{EPs generated by two interacting fermions in a system of length $L=3$ (left) and $L=6$ (right).
    The color code in panel a) and b) indicates the minimal angle enclosed by two eigenvectors for each point in the parameter space spanned by the interaction strength $U$ and twist angle $\varphi$. Dark blue lines indicate the path of EPs which are inherited from the single particle spectrum (i), $\varphi_{\text{e}}$, or emerge from a diagonalizable degeneracy at $U=0$ (ii), $\varphi_\text{d}$, and form an endpoint at (iii), $\varphi_\text{a}$ and $U_\text{a}\neq 0$. We evaluate the minimal angle on the circles (parameterized by $\theta$) drawn in panel a) (b)) and show the results on a semilogarithmic scale in panels c) (d)) for $m=0.7$. The point $P$ refers to Fig. \ref{fig:EP_vector_app}.}
\end{figure}

The starting point of our perturbative ansatz are eigenstates of the two-particle Hamiltonian in the non-interacting limit, Eq. \eqref{eq:H0_model_2_fermions}. 
Similar to perturbative treatments in the hermitian case we generate an effective Hamiltonian based on states which have the same eigenvalue for $U = 0$. The effective matrix is generated from the corresponding right and left (generalized) eigenvectors. By assuming the effective Hamiltonian to be defective, we can determine conditions for $U$ predicting the paths of EPs in the parameter space.
\begin{equation}
	H^{\text{eff}} = \sum_{ij}h_{ij}\ket{\Psi_i^R}\bra{\Psi_j^L}\text{ with } h_{ij} = \braOket{\Psi_i^L}{H}{\Psi_j^R}
\end{equation}
First, we can identify trivial eigenstates from Eq. \eqref{eq:H0_model_2_fermions} which are defined for a single momentum in the second term of the two-particle Hamiltonian exhibiting the eigenvalue zero:
\begin{align}
	\ket{\Phi_k}=&\ket{ E_{(k,+)};E_{(k,-)}} = a_k^\dagger b_k^\dagger \ket{0}\\
	\text { and }	\bra{\Phi_k}=&\bra{E_{(k,+)};E_{(k,-)} } = \bra{0} b_k a_k\nonumber
\end{align}
Second, away from an EP, the remaining states can be derived from the $4\times 4$ matrix and refer to the four possible eigenenergies $E_{(k,\xi_k)} + E_{(q,\xi_q)}$ with $\xi_k,\xi_q=\pm$. They are constructed from \textit{Fourier}-states which are contained in $\vec{\Psi}_{k,q}^\dagger$.
\begin{align}
	\ket{ E_{(k,\xi_k)};E_{(q,\xi_q)}} =& \left(\prod_{d=k,q}\frac{ \left(\xi_d\sqrt{m_d}a_d^\dagger + \sqrt{p_d} b_d^\dagger\right) }{\sqrt{2E_{(d,+)}}}\right)\ket{0}\label{eq:EV_k0_k1}\\
	\bra{E_{(k,\xi_k)};E_{(q,\xi_q)}} =& \bra{0}\left(\prod_{d=q,k}\frac{ \left(\xi_d\sqrt{p_d}a_d+ \sqrt{m_d} b_d\right) }{\sqrt{2E_{(d,+)}}}\right)\label{eq:LEV_k0_k1}
\end{align}
Note that the positions of $k$ and $q$ are swapped for the left and right eigenvector. The eigenstates fulfill the orthogonality relation such that $\braket{E_{(k,\tilde{\xi}_{k})};E_{(q,\tilde{\xi}_q)}}{ E_{(k,\xi_k)};E_{(q,\xi_q)}} = \delta_{\tilde{\xi}_k\xi_k} \delta_{\tilde{\xi}_q\xi_q}$. Two states exhibiting two different momenta are orthogonal due to the block structure of the Hamiltonian.

\subsection{(i) Inherited EPs}
\begin{figure}
	\centering
	\includegraphics[width=\columnwidth]{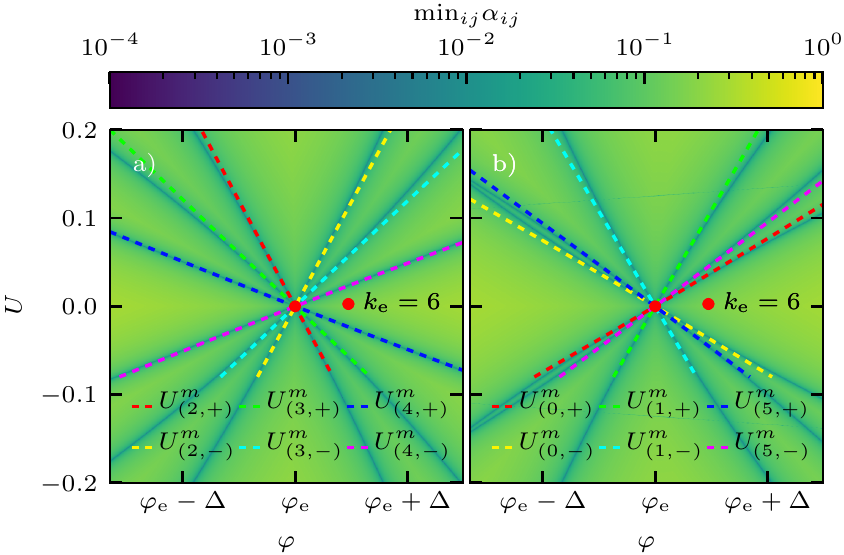}\label{fig:EP_stability}
	\caption{Spreading of exceptional lines for $L=7$ and $m=0.6$ with a hermitian interaction ($U\in\mathbb{R}$) in panel a) and an anti-hermitian interaction ($U\in i\mathbb{R}$) in the panel b). The EP located in the single particle spectrum is obtained for $k_{\text{e}} = 6$ and $\varphi_{\text{e}}=2 (\pi + \arctan((5 - \sqrt{41})/8)) L - 2\pi k_{\text{e}}$, $\Delta=0.05$. The predicted trajectories of EPs are plotted as dashed lines according to the interaction strength in Eq. \eqref{eq:eps_stability_app}.}
\end{figure}

As a first source of EPs in the many-body case, we find that the defective structure is inherited from the single particle spectrum which is located at $k_{\text{e}}$ and $\varphi_{\text{e}}$. The two-band Bloch Hamiltonian becomes defective at $\varphi_{\text{e}}$ if $m_{k_{\text{e}}}=0$ or $p_{k_{\text{e}}}=0$ and transforms into a $2\times 2$ Jordan block. Since the Jordan block is non-diagonalizable we generate the effective Hamiltonian from the generalized eigenvectors $a_{k_{\text{e}}}^\dagger\ket{0}$ and $b_{k_{\text{e}}}^\dagger\ket{0}$ and an additional single particle state $\ket{E_{(q,\pm)}}$:
\begin{align}
	\ket{c_{k_{\text{e}}};E_{(q,\pm)}} &= \left(\frac{\pm\sqrt{m_q}c_{k_{\text{e}}}^\dagger a_q^\dagger + \sqrt{p_q}c_{k_{\text{e}}}^\dagger b_q^\dagger}{\sqrt{2E_{(q,+)}}}\right)\ket{0}\\
		\bra{c_{k_{\text{e}}};E_{(q,\pm)}} &= \bra{0}\left(\frac{\pm\sqrt{p_q}a_q c_{k_{\text{e}}}  + \sqrt{m_q}b_q c_{k_{\text{e}}} }{\sqrt{2E_{(q,+)}}}\right)
\end{align}
This choice of generalized left and right eigenvectors obeys $\braket {c_{k_{\text{e}}};E_{(q,\pm)}}{ \tilde{c}_{k_{\text{e}}};E_{(q,\pm)}} = \delta_{c,\tilde{c}}$. The matrix elements are given by ($c=a,b$)
\begin{align}
	\braOket{ c_{k_\text{e}};E_{(q,\pm)}}{ H_{\text{int}} }{\tilde{c}_{k_\text{e}};E_{(q,\pm)}} & = \frac{1}{2L}\delta_{c\tilde{c}}\label{eq:Hint0}\\
	\braOket{a_{k_{\text{e}}};E_{(q,\pm)}}{ H_{\text{int}} }{ b_{k_{\text{e}}};E_{(q,\pm)}} & = \mp\frac{1}{2L}\frac{\sqrt{m_q}}{\sqrt{p_q}}\label{eq:Hint1}\\
	\braOket{b_{k_{\text{e}}};E_{(q,\pm)}}{ H_{\text{int}} }{ a_{k_{\text{e}}};E_{(q,\pm)}} & = \mp\frac{1}{2L}\frac{\sqrt{p_q}}{\sqrt{m_q}}\label{eq:Hint2}
\end{align}
The full effective Hamiltonian of size $2\times 2$ which is spanned by $\vert a_{k_{\text{e}}};E_{(q,\pm)}\rangle$ and $\vert b_{k_{\text{e}}};E_{(q,\pm)}\rangle$ is:
\begin{align}
	H^{\text{(i)}} = \begin{pmatrix} E_{(q,\pm)} & m_{k_{\text{e}}} \\ p_{k_{\text{e}}} & E_{(q,\pm)}\end{pmatrix} +  \frac{U}{2L}\begin{pmatrix} 1 &  \mp\sqrt{\nicefrac[\texttt]{$m_q$}{$p_q$}} \\ \mp \sqrt{\nicefrac[\texttt]{$p_q$}{$m_q$}} & 1 \end{pmatrix}\label{eq:eff_Ham_stability_app}
\end{align}
Tuning the effective Hamiltonian to $\varphi=\varphi_{\text{e}}$ and $U = 0$ reveals the defective structure since $m_{k_{\text{e}}}=0$ or $p_{k_{\text{e}}}=0$. Now we can derive conditions for $U(\varphi)$ which preserves the Jordan block of the effective matrix for $\varphi \neq \varphi_{\text{e}}$. The diagonal elements remain equal for finite $U$ such that the matrix is defective if and only if $H^{\text{(i)}}_{01}=0$ or $H^{\text{(i)}}_{10}=0$ which induces
\begin{align}
	U_{(q,\pm)}^m =\pm{2L}{m_{k_{\text{e}}}}\frac{\sqrt{p_q}}{\sqrt{m_q}}\text{ and }U_{(q,\pm)}^p =\pm{2L}{p_{k_{\text{e}}}}\frac{\sqrt{m_q}}{\sqrt{p_q}}\label{eq:eps_stability_app}
\end{align}
The derived formulas predict the paths $U(\varphi)$ of EPs emerging from the Jordan block at $\varphi=\varphi_{\text{e}}$ and $U=0$. It can be used to evaluate the existence of the EPs for a finite interaction strength. As long as the system does not undergo an other transition via an EP in the single particle spectrum the solutions for $U$ are either real or imaginary. If the energy $E_{(q,\pm)}$ is real (imaginary), the prediction of $U$ is real (imaginary). Therefore, the EP which is formed in the non-interacting limit will be either present for a hermitian interaction, $U \in \mathbb{R}$, or an anti-hermitian interaction, $U\in i \mathbb{R}$. Fig. \ref{fig:EP_stability} shows the characteristic ''fan'' emerging from the same EP for a hermitian (panel a)) and anti-hermitian (panel b)) interaction and compares it to the prediction via Eq. \eqref{eq:eps_stability_app}.
Furthermore, the effective Hamiltonian allows us to identify the eigenstate which forms the EP. Starting from the non-interacting limit and $\varphi=\varphi_{\text{e}}$ the exceptional state will remain in $\ket{a_{k_{\text{e}}},E_{(q,\pm)}}$ or $\ket{b_{k_{\text{e}}},E_{(q,\pm)}}$ and is stable as long as the perturbative approach is valid.

\subsection{(ii) Emergent EPs}
While EPs are induced from a non-diagonalizable matrix in the single particle spectrum in the previous section, we find a second source which is limited to the case of interacting particles, $U\neq 0$. The model is purely diagonalizable in the non-interacting limit (for $\varphi\neq \varphi_{\text{e}}$). However, we demonstrate how EPs can emerge from a diagonalizable degeneracy for a finite interaction strength. Two particle eigenstates are generated from two single particle states with energy $E_{(k,\xi_k)}$ and $ E_{(q,\xi_q)}$ where $\xi_k,\xi_q = \pm $. Combining two different momenta ($k\neq q$) yields four different two-particle states exhibiting the energies $E_{(k,\pm)} + E_{(q,\pm)}$. The corresponding left and right eigenstates are defined in Eq. \eqref{eq:EV_k0_k1} and Eq. \eqref{eq:LEV_k0_k1}.
We need to evaluate the matrix elements of the density-density interaction to generate the effective Hamiltonian:
 \begin{align}
	&\braOket{ E_{(k_0,\xi_0)};E_{(k_1,\xi_1)}}{ H_{\text{int}} }{E_{(k_2,\xi_2)};E_{(k_3,\xi_3)} }\label{eq:perturbation_general} \\
=&\delta_{k_0+k_1,k_2+k_3}\frac{1}{4L\sqrt{E_{(k_0,+)}E_{(k_1,+)}E_{(k_2,+)}E_{(k_3,+)}}} \nonumber\\
	&\left(\xi_1\xi_3\sqrt{m_{k_0}p_{k_1}p_{k_2}m_{k_3}}-\xi_1\xi_2\sqrt{m_{k_0}p_{k_1}m_{k_2}p_{k_3}}\right.\nonumber\\
	&\,\left.\xi_0\xi_2\sqrt{p_{k_0}m_{k_1}m_{k_2}p_{k_3}}-\xi_0\xi_3\sqrt{p_{k_0}m_{k_1}p_{k_2}m_{k_3}}\right)\nonumber
\end{align}
The most common source for degeneracies in the two-particle spectrum are induced from degeneracies in the single particle spectrum for different momenta, $E_{(k,+)}=E_{(q,-\xi)}$, ($k\neq q$ and $\xi=\pm$) at $\varphi=\varphi_\text{d}$. Two eigenenergies of the two-body Hamiltonian
\begin{align}
	\pm\delta =  E_{(k,\pm)} +E_{(q,\pm\xi)}
\end{align}
coalesce with $\pm\delta =0$ at $\varphi=\varphi_\text{d}$. Even though the system has degenerated eigenvalues it exhibits distinct eigenvectors as defined in Eq. \eqref{eq:EV_k0_k1} and \eqref{eq:LEV_k0_k1}. We define the states referring to the energy $+\delta$ with $\ket{ \Psi_+}$ and $-\delta$ with $\ket{ \Psi_-}$.

Again, we construct an effective model including  all states with the same eigenvalue, here zero, and the same total momentum for $U=0$. The non-interacting model naturally exhibits states with eigenvalue zero: $\ket{\Phi_{p}} = \ket{E_{(p,+)};E_{(p,-)}}=a_p^\dagger b_p^\dagger\ket{0}$. Therefore, we need to include the additional states (or their superposition) in our effective description if the total momentum agrees: $k+q=2p$. Whether or not such a state exists in the correct momentum sector depends on the system size and total momentum. We obtain an effective Hamiltonian of size $2\times 2$ (even $L$, odd $k+q$),  $3\times 3$ (odd $L$) or $4\times 4$ (even $L$, even $k+q$). The $3\times 3$ matrix is extended by a single state $\vert \Phi_{{p}}\rangle$ and the $4\times 4$ matrix is extended by $\vert\Phi_\pm\rangle = \vert \Phi_{{p}}\rangle/\sqrt{2} \pm \vert \Phi_{{p}^\prime}\rangle/\sqrt{2}$ with $2{p}=2{p}^\prime=k+q$ ($\vert \Phi_-\rangle$ is the first state in the $4\times 4$ matrix): 
\begin{align}
	H^{\text{(ii)}}_{2\times 2} &= \begin{pmatrix}  \delta & 0 \\  0 & -\delta \end{pmatrix} +  \frac{U}{L}\begin{pmatrix} -\xi a^2 &  \xi a^2 \\  \xi a^2 & -\xi a^2 \end{pmatrix}\label{eq:eff_Ham_ii}\\
	H^{\text{(ii)}}_{3\times 3} &= \begin{pmatrix}  0 & 0 & 0 & \\ 0 & \delta & 0 \\  0 & 0 & -\delta \end{pmatrix} +  \frac{U}{L}\begin{pmatrix} 1 & \xi a & -\xi a \\ -a & -\xi a^2 & \xi a^2 \\ a &  \xi a^2 & -\xi a^2 \end{pmatrix}\nonumber\\
	H^{\text{(ii)}}_{4\times 4} &= \begin{pmatrix}  0 & 0 & 0 & 0 & \\ 0 & 0 & 0 & 0 & \\ 0 & 0 & \delta & 0 \\  0 & 0 & 0 & -\delta \end{pmatrix} +  \frac{U}{L}\begin{pmatrix} 0 & 0 & 0 & 0 \\ 0 & 2 &  \xi\sqrt{2} a & - \xi \sqrt{2} a \\  0 &-\sqrt{2}a & -\xi a^2 & \xi  a^2 \\ 0 & \sqrt{2}a & \xi a^2 & -\xi a^2 \end{pmatrix}\nonumber\\
	\text{with } a &=\frac{\sqrt{m_{k}p_{q}}-\xi \sqrt{p_{k}m_{q}} }{2\sqrt{E_{(k,+)}}\sqrt{E_{(q,+)}}}
\end{align}
Until the system undergoes a transition via an EP in the single particle spectrum, both included single particle energies ($E_{(k,+)}$ and $E_{(q,+)}$) are either real or imaginary. This induces $\delta$ to be purely real or imaginary. Also, $a$ is either real or imaginary yielding $a^2\in\mathbb{R}$. We derive constraints for $U$ such that the effective Hamiltonian is defective. The case of the $2\times 2$ matrix is particularly simple and yields
\begin{align}
	U^\pm_{2\times 2} = \pm i\frac{L\delta}{a^2}\label{eq:EP_2x2_app}.
\end{align}

\begin{figure}
	\centering
	\includegraphics[width=\columnwidth]{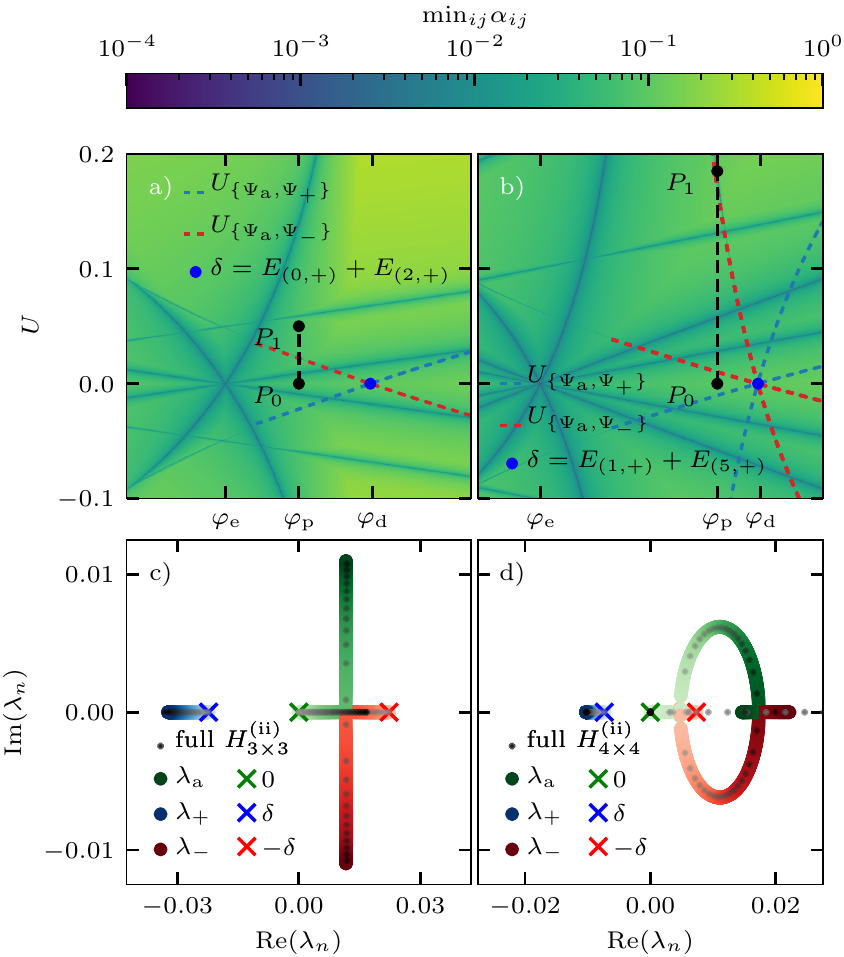}\label{fig:EP_emergence}
	\caption{We are analyzing the non-trivial eigenvalues of the effective Hamiltonian $H^{\text{(ii)}}_{3\times 3}$ (left panels) and $H^{\text{(ii)}}_{4\times 4}$ (right panels). The lines of emergent EPs (dashed blue/red) are predicted via Eq. \eqref{eq:EP_3x3_app} and Eq. \eqref{eq:EP_4x4_app}. The left (right) panels are showing a system with $L=3$ ($L=6$) sites for $m=0.7$ and are modeled by an effective Hamiltonian of size $3\times 3$ ($4\times 4$). We can identify the analytical eigenvalues (Eq. \eqref{eq:eigenvalues}) with $\lambda_\text{a}=0$, $\lambda_+=\delta$ and $\lambda_-=-\delta$ and the corresponding eigenstates $\ket{\Psi_\text{a}}$, $\ket{\Psi_+}$ and $\ket{\Psi_-}$. Starting from $U=0$, we can track the eigenvalues continuously using Riemann surfaces as shown in the lower panels. We track the eigenvalues from $P_0$ at $\varphi_\text{p}$ and $U=0$ to a finite $U$ at $P_1$. The initial eigenvalues are marked by crosses and the color shading from light to dark indicates the transition from $P_0\rightarrow P_1$ in panels c) and d). We find that the EP (red dashed line) is formed by states which are initially associated with $\lambda_\text{a}=0$ and $\lambda_-=-\delta$. While the left panel reveals only one intersection of the eigenvalues, the right panel shows two intersections. The second solution of Eq. \eqref{eq:EP_3x3_app} is imaginary and therefore, only one is present in the case of a hermitian interaction. The gray dots in the lower panels are showing the eigenvalues of the full Hamiltonian evaluated from $P_0\rightarrow P_1$ which agree remarkable well with the effective description.}
\end{figure}

The EP is only present in the $U$--$\varphi$ plane for a hermitian interaction if $\delta\in i\mathbb{R}$ and only exists for an anti-hermitian interaction if $\delta\in\mathbb{R}$.

The $3\times 3$ and $4\times 4$ is hermitian for $a\in i\mathbb{R}$, $\xi = 1$ and $a\in \mathbb{R}$, $\xi = -1$. Therefore, the effective Hamiltonian does not exhibit EPs for $U\in\mathbb{R}$ in these cases.
However, apart from these cases, we can derive constraints for $U$ such that the effective matrix becomes defective. The eigenvalues are given by

\begin{align}
	\lambda_{n} =& -\frac{e^{-i2\pi n/3}}{3} \frac{x}{\sqrt[3]{\sqrt[2]{y_{m\times m}^2-x_{m\times m}^3}+y_{m\times m}}}\label{eq:eigenvalues}\\
	&-\frac{e^{i2\pi n/3}}{3}\sqrt[3]{\sqrt[2]{y_{m\times m}^2-x_{m\times m}^3}+y_{m\times m}}+c_{m\times m}\nonumber 
\end{align}
for $n=0,1,2$ and $m = 3,4$. The forth eigenvalue $\lambda_3 = 0$ of $H^{\text{(ii)}}_{4\times 4}$ is trivial and does not form an EP. A hermitian interaction, $U\in\mathbb{R}$, induces $c_{m\times m}$, $x_{m\times m}$ and $y_{m\times m}$ to be real numbers.

\begin{align}
	x_{3\times 3} =& 3\delta^2 +\frac{U^2(2a^2-\xi)^2}{L^2}\\
	x_{4\times 4} =& 3\delta^2 +\frac{U^2(2a^2-2\xi)^2}{L^2}\nonumber\\
	y_{3\times 3} =& \xi\frac{9\delta^2U(a^2+\xi)}{L}+\xi\frac{U^3(2a^2-\xi)^3}{L^3}\nonumber\\
	y_{4\times 4} =& \xi \frac{9\delta^2U(a^2+2\xi)}{L}+\xi\frac{U^3(2a^2-\xi 2)^3}{L^3}\nonumber\\
	c_{3\times 3} =& \xi U\frac{2\xi - 2a^2}{3L},\quad c_{4\times 4} = \xi U\frac{2\xi -2a^2}{3L}\nonumber
\end{align}

EPs are formed if two eigenvalues coincide. Setting the difference of any two eigenvalues in Eq. \eqref{eq:eigenvalues} to zero yields $x_{m\times m}^3=y_{m\times m}^2$ and induces four independent solutions for $U$.

\begin{align}
	U^{\pm,\pm}_{3\times 3} = \pm L\delta\sqrt{\xi\frac{a^4 -\xi 10 a^2 - 2 \pm a\sqrt{(a^2+4\xi )^3}}{2(2a^2-\xi)^3}}\label{eq:EP_3x3_app}\\
	U^{\pm,\pm}_{4\times 4} = \pm L\delta\sqrt{\xi\frac{a^4 -\xi 20 a^2 - 8 \pm a\sqrt{(a^2+8\xi )^3}}{32(a^2-\xi)^3}}\label{eq:EP_4x4_app}
\end{align}
Again, we can use the derived constrains for $U$ in Eq. \eqref{eq:EP_3x3_app} and Eq. \eqref{eq:EP_4x4_app} to evaluate the stability for a finite hermitian interaction. If any solution of $U$ is purely real, it will spread within for $U$--$\varphi$ plane starting from $\varphi_\text{d}$. While earlier solutions for $U$ are either real or imaginary, $U$ can be a complex number and is not restricted to propagate within the purely hermitian or purely anti-hermitian case.

Furthermore, the analytic approach allows us to assign the corresponding eigenstates to the eigenvalues which form the EP. Starting from the non-interacting limit, we can associate the eigenvalues $\lambda_{\text{a}} = 0$, $\lambda_{+} = \delta$ and $\lambda_{-} = -\delta$ to $\ket{\Psi_\text{a}}$, $\ket{ \Psi_+} = \ket{ E_{(k,+)}; E_{(q,\xi)}}$ and $\ket{ \Psi_-}= \ket{ E_{(k,-)}; E_{(q,-\xi)}}$ respectively. Here, $\ket{ \Psi_\text{a}}$ refers to $a_{p}^\dagger b_{p}^\dagger \ket{0}$ in the case of the $3\times 3$ matrix and $\left(a_{p}^\dagger b_{p}^\dagger+a_{p^\prime}^\dagger b_{p^\prime}^\dagger\right)/\sqrt{2}\ket{0} $ in the case of the $4\times 4$ matrix. First, we identify the correct eigenvalues away from the degeneracy in the non-interacting limit at $U=0$ and $\varphi_\text{p}\neq\varphi_\text{d}$. Second, we adiabatically track the eigenvalues from the non-interacting limit to the EP, $\left(\varphi_\text{p},0\right)\rightarrow \left(\varphi_\text{p},U^{\pm,\pm}_{m\times m}\right)$, using Riemann surfaces which is necessary since the roots appearing in the expressions for the eigenvalues are not defined uniquely. The procedure is illustrated in Fig. \ref{fig:EP_emergence}. The different colors indicate the two states which form the EP in panel a) and b).

More generally, non-zero degeneracies can occur in the case of two fermions. Two states given by $\ket{\Psi_{0,+}} = \ket{ E_{(k,\xi_k)}; E_{(q,\xi_q)}}$ and $\ket{ \Psi_{1,+}} = \ket{ E_{(p,\xi_p)}; E_{(n,\xi_n)}}$ can have the same energy $\delta=E_{(k,\xi_k)}+E_{(q,\xi_q)}=E_{(p_,\xi_p)}+E_{(n,\xi_n)}$ for $\varphi=\varphi_\text{d}$ and form an EP in the interacting case if $k+q=p+n$. This induces that the states $\ket{\Psi_{0,-}} = \ket{E_{(k,-\xi_k)}; E_{(q,-\xi_q)}}$ and $\ket{ \Psi_{1,-}} = \ket{E_{(p,-\xi_p)}; E_{(n,-\xi_n)}}$ are degenerated with the energy $-\delta$. Again, we can construct an effective Hamiltonian of size $2\times 2 $ and derive constraints for $U$.
Also it should be mentioned that the system incorporates high symmetry points at $\varphi=0,\pi$ where degeneracies occur naturally and EPs are emerging.

\begin{figure}
	\centering
	\includegraphics[width=\columnwidth]{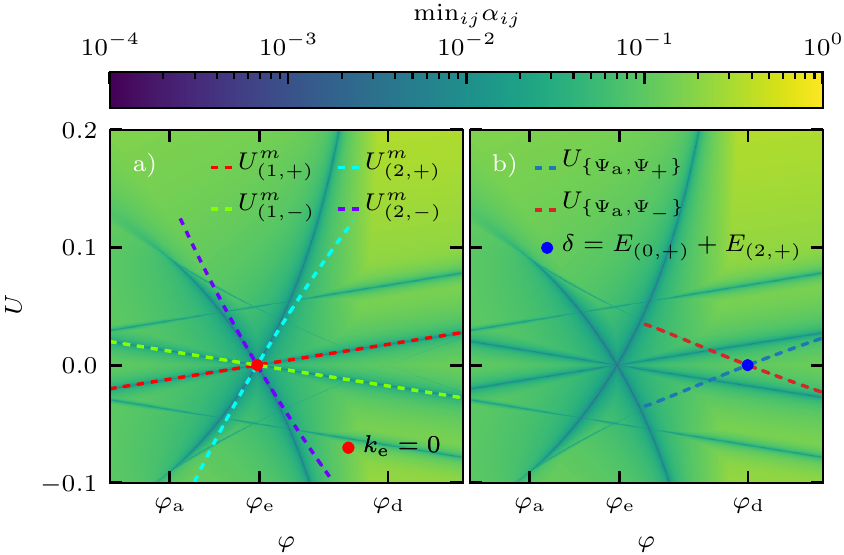}\label{fig:EP_prediction_app}
	\caption{Comparison of our analytic predictions (dashed lines) of EPs to the numerical simulation for $L=3$ and $m=0.7$. We identify three different twist angles where (i) EPs are inherited from the single particle spectrum at $\varphi_{\text{e}}$, (ii) emerge from a diagonalizable degeneracy at $\varphi_\text{d}$ and (iii) annihilate each other at $\varphi_\text{a}$. Panel a) and b) show the analytical predictions from Eq. \eqref{eq:eps_stability_app} and \eqref{eq:EP_3x3_app} and identify the states forming the exceptional lines.}
\end{figure}

Fig. \ref{fig:EP_prediction_app} compares the prediction of our perturbative treatment for a system of $L=3$ sites with numerical simulations and finds an excellent agreement.

\subsection{(iii) Annihilation}

\begin{figure}
	\centering
	\includegraphics[width=\columnwidth]{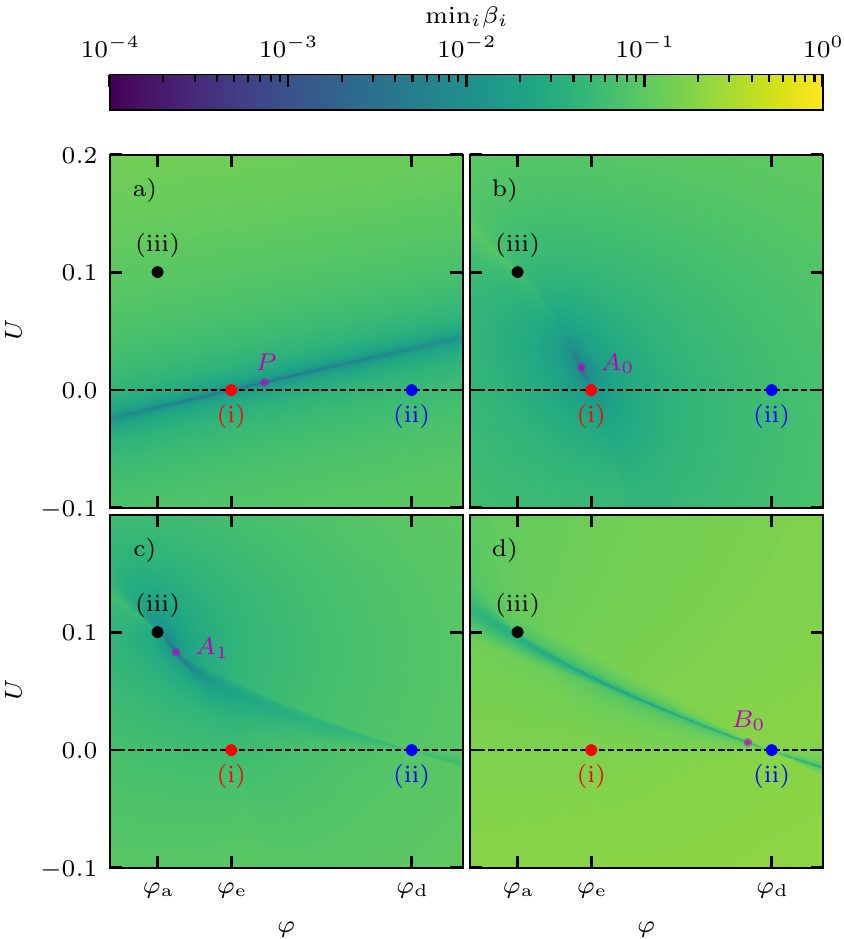}\label{fig:EP_vector_app}
	\caption{Robustness of the eigenvector associated with the lines of EPs for a system of $L=6$ sites and $m=0.7$ (\cf Fig. \ref{fig:circles_app}). Each reference point ($P$, $A_0$, $A_1$ and $B_0$) refers to an EP which can be found in Fig. \ref{fig:circles_app}. We determine the eigenvector $\ket{\Psi^R_{\text{EP}}}$ for the reference point and calculate the minimal angle between all eigenstates for each point in the parameter space spanned by the twist angle $\varphi$ and interaction strength $U$. The quantifier is given by $\text{min}_i\beta_{i} = \text{min}_i\arccos\left(\vert\braket{\Psi_i^R}{\Psi^R_{\text{EP}}}\vert\right)$.}
\end{figure}

Besides the emergence of exceptional lines we also find their annihilation at finite interaction strength. Some lines of EPs are forming an endpoint while others simply cross in the parameter space. A first indicator is the conserved total momentum which protects exceptional lines emerging in different momentum sectors. However, a second mechanism must be present to allow some EPs in the same momentum sector to interact and essentially form an endpoint while others do not. 

To better understand this phenomenon we can evaluate the aligned eigenvector forming the characteristic exceptional ``fans''. The excellent agreement of our perturbative treatment with the numerical simulation suggests that our approach describes not only the paths but also the corresponding eigenvectors. EPs inherited from the single particle spectrum (i) are described by the effective Hamiltonian in Eq. \eqref{eq:eff_Ham_stability_app}. The defective structure of the matrix is realized if one off-diagonal element is zero yielding to Eq. \eqref{eq:eps_stability_app}. Starting from the non-interacting limit and keeping the corresponding off-diagonal matrix zero yields the same eigenvector $\ket{a_{k_{\text{e}}},E_{(q,\pm)}}$ or $\ket{b_{k_{\text{e}}},E_{(q,\pm)}}$ for different twist angles $\varphi$ and interaction strengths $U\neq 0$ away from the critical point (i).


Fig. \ref{fig:EP_vector_app} is evaluating the robustness of the eigenvector forming the lines of EPs starting from a reference point ($P$, $A_0$, $A_1$ and $B_0$ in Fig. \ref{fig:circles_app}b)). While EPs emerging from diagonalizable degeneracies (ii) are robust and the eigenvector remains in the same state (as long as it is not annihilated, \cf panel d)), EPs inherited from the single particle spectrum (i) exhibit two scenarios. First, eigenvectors which are \textit{not} annihilated remain in the same state (\cf panel a)) throughout the parameter space. Second, exceptional states which form an endpoint with another EP continuously transforms themselves into their annihilation partner (\cf panel b)). Initially, the state emerging from the single particle spectrum (i) is given by $\ket{a_{k_{\text{e}}}E_{(q,\pm)}}$ (or $\ket{b_{k_{\text{e}}}E_{(q,\pm)}}$). However, as indicted in panel c) the exceptional state at $A_1$ (which is connected to (i), \cf Fig. \ref{fig:circles_app}b)) is given by the exceptional state emerging from (ii). 

\begin{table}[t]
	\centering
		\begin{tabular}{c | c | c }
			 & $L=3$, Fig. \ref{fig:EP_prediction_app} & $L=6$, Fig. \ref{fig:EP_prediction} \\ [0.5ex]
			\hline
			(i) line A  & $\ket{a_0,E_{(2,-)}}$   & $\ket{a_1,E_{(5,-)}}$ \\
			(ii) line B & $\ket{E_{(0,-)},E_{(2,-)}}$ & $\ket{E_{(1,-)},E_{(5,-)}}$ \\
			\end{tabular}
			\caption{The table lists the involved two-body states forming the lines of EPs (A and B) in Fig. \ref{fig:circles_app} for a system of $L=3$ and $L=6$ sites which form an endpoint (iii). The involved states can be extracted from Fig. \ref{fig:EP_prediction} (main text) and Fig. \ref{fig:EP_prediction_app}.}
		\label{tab:states}
\end{table}
\begin{figure}
	\centering
	\includegraphics[width=\columnwidth]{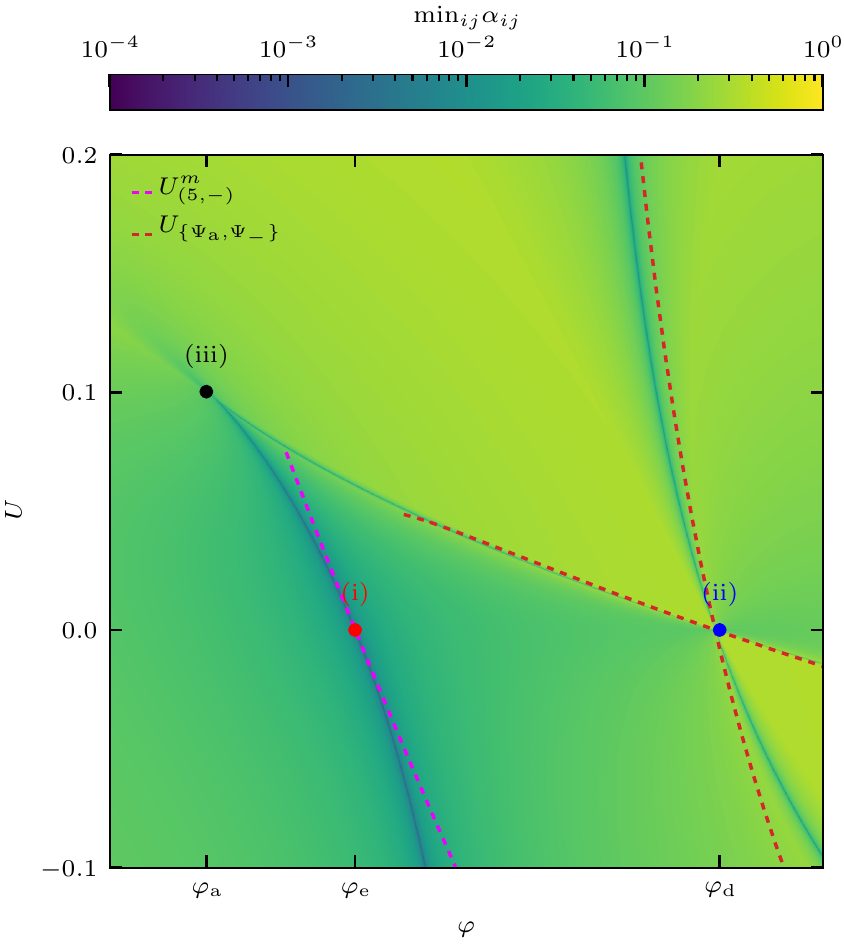}\label{fig:eff_H_iii}
	\caption{Numerical evaluation of the effective Hamiltonian $H^{\text{(iii)}}_{\text{even}}$ in Eq. \eqref{eq:H_iii_even} where we determine the minimal angle between its eigenstates. The perturbative description captures all phenomena including the heredity (i), emergence (ii) and annihilation (iii) of EPs in a system with $L=6$ sites and $m=0.7$. The corresponding analytical predictions $U^m_{(q,\pm)}$ in Eq. \eqref{eq:eps_stability_app} derived from $H^{\text{(i)}}$ and $U_{\{\Psi_\text{a},\Psi_-\}}$ in Eq. \eqref{eq:EP_4x4_app} derived from $H^{\text{(ii)}}_{4\times 4}$ agree with the numerical calculation.}
\end{figure}

Furthermore, our perturbative treatment allows us to identify the single particle states which are forming the EPs using Eq. \eqref{eq:eps_stability_app} and Eq. (\ref{eq:EP_2x2_app}, \ref{eq:EP_3x3_app}, \ref{eq:EP_4x4_app}). We observe that two lines form an endpoint if they are initially generated from the same single particle states. 
Fig. \ref{fig:circles_app} marks two lines ($A_0\rightarrow A_1$ and $B_0\rightarrow B_1$) which form an endpoint for a system of $L=3$ and $L=6$ sites. While the exceptional line $A$ is inherited from the single particle spectrum (i), line $B$ emerges from diagonalizable degeneracy (ii). The involved states are listed in Tab. \ref{tab:states}. Both lines include the single particle states $\ket{E_{(2,-)}}$ and $\ket{a_0}$ ($\ket{E_{(5,-)}}$ and $\ket{a_1}$) in the case of $L=3$ ($L=6$) sites. 

Our perturbative prediction from $H^{\text{(i)}}$, Eq. \eqref{eq:eff_Ham_stability_app}, is more robust for the inherited EPs that are \textit{not} annihilated. This suggests that the perturbative subspace is insufficient in this case and needs to be extended to capture the whole phenomenology including the annihilation process. A complete perturbative description requires the two generalized eigenvectors forming the effective $2\times 2$ Hamiltonian $H^{\text{(i)}}$ at $k_\text{e}$ and both states which we identified in the EP emerging from (ii). Both EPs include the same diagonalizable single particle state $\ket{E_{(q,\xi)}}$, $\xi=\pm$, which is combined with the single particle EP located at $k_\text{e}$ to form two-particle states $\ket{a_{k_\text{e}};E_{(q,\xi)}}$ and $\ket{b_{k_\text{e}};E_{(q,\xi)}}$. Next to the additional state $\ket{\Psi_\text{a}}$, the second two-particle state necessary for the emergence of the EP in (ii) is $\ket{\Psi_\pm} = \ket{E_{(k_\text{e},\pm)};E_{(q,\pm \xi)}}$. However, we can omit $\ket{\Psi_\pm}$ in our perturbative subspace as it includes the single particle states $\ket{E_{(q,\pm)}}$ and $\ket{E_{(k_\text{e},\pm)}}$ (due to the conservation of the total momentum) and is linearly depending on $\ket{a_{k_\text{e}};E_{(q,\pm)}}$ and $\ket{b_{k_\text{e}};E_{(q,\pm)}}$. Therefore, a complete perturbative description can be obtained by extending the $2\times 2$ effective Hamiltonian $H^{\text{(i)}}$ by the additional state $\ket{\Psi_\text{a}}$. Depending on the additional state which is given by a single trivial state (odd $L$) or a superposition of two trivial states (even $L$ and even $k_\text{e}+q$) we obtain two different effective Hamiltonians of size $3 \times 3$ labeled by $H^{\text{(iii)}}_\text{odd}$ and $H^{\text{(iii)}}_\text{even}$:
\begin{align}
	&H^{\text{(iii)}}_{\text{odd}} = \begin{pmatrix}  0 & 0 & 0 & \\ 0 & E_{(q,\pm)} & m_{k_\text{e}} \\  0 & p_{k_\text{e}} & E_{(q,\pm)}\end{pmatrix} \label{eq:H_iii_odd}\\
	&+  \frac{U}{2L}\begin{pmatrix} 2 & \sqrt[4]{4p_q/m_q} & \mp\sqrt[4]{4m_q/p_q}\\  \sqrt[4]{4m_q/p_q} & 1 & \mp \sqrt{m_q/p_q} \\  \mp\sqrt[4]{4p_q/m_q} & \mp \sqrt{p_q/m_q} & 1 \end{pmatrix}\nonumber\\
	&H^{\text{(iii)}}_{\text{even}} = \begin{pmatrix}  0 & 0 & 0 & \\ 0 & E_{(q,\pm)} & m_{k_\text{e}} \\  0 & p_{k_\text{e}} & E_{(q,\pm)}\end{pmatrix} \label{eq:H_iii_even} \\
	&+  \frac{U}{2L}\begin{pmatrix} 4 & \sqrt[4]{16p_q/m_q} & \mp\sqrt[4]{16m_q/p_q}\\  \sqrt[4]{16m_q/p_q} & 1 & \mp \sqrt{m_q/p_q} \\  \mp\sqrt[4]{16p_q/m_q} & \mp \sqrt{p_q/m_q} & 1 \end{pmatrix}\nonumber
\end{align}
The matrices are derived from the left and right eigenvectors associated to the states $\ket{\Psi_\text{a}}$, $\ket{a_{k_\text{e}};E_{(q,\pm)}}$ and $\ket{b_{k_\text{e}};E_{(q,\pm)}}$. Fig. \ref{fig:eff_H_iii} evaluates minimal angles between the three eigenvectors of the effective Hamiltonian for given $\varphi$ and $U$ and finds an excellent agreement between the effective and full Hamiltonian. The extended description captures all phenomena including the heredity (i), emergence (ii) and annihilation (iii) of EPs. We carefully examined the annihilation point (iii) numerically and find a \emph{third} order EP matching with the three dimensional perturbative subspace.




\begin{figure}
	\centering
	\includegraphics[width=\columnwidth]{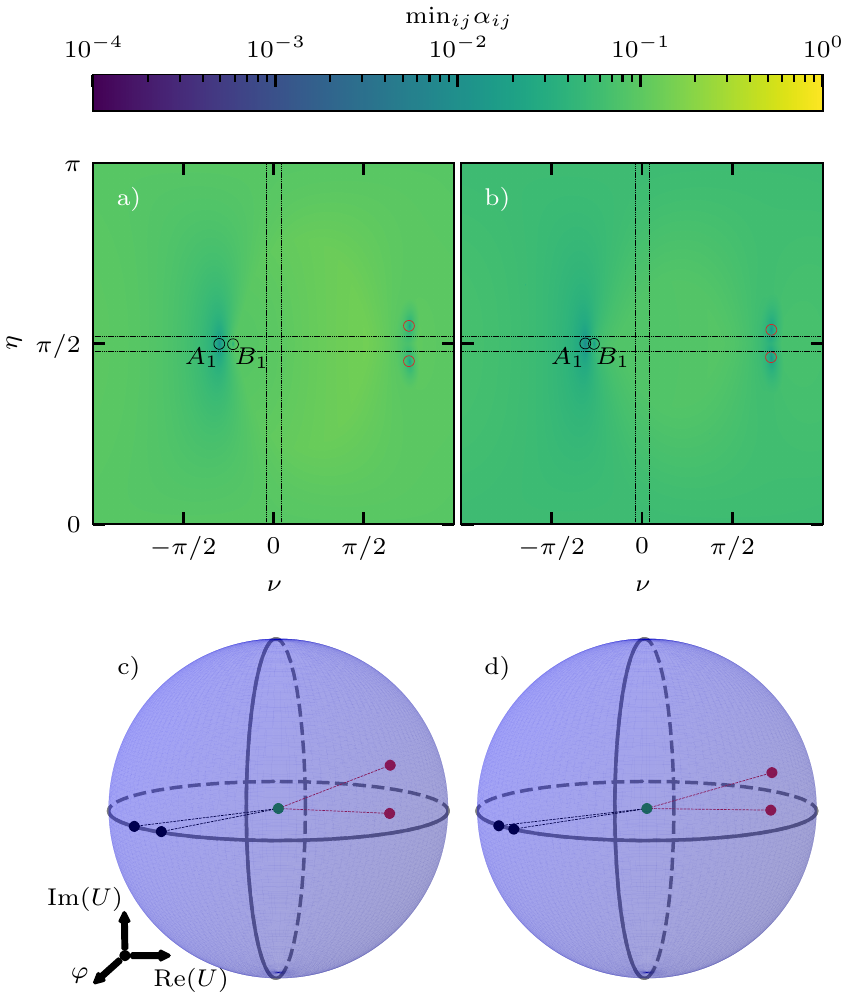}\label{fig:sphere_app}
	\caption{Panel a) ($L=3$) and b) $(L=6)$ are showing the minimal angle enclosed by all eigenvectors of a system parameterized on a sphere around the endpoints (iii) in Fig. \ref{fig:circles_app} which is extended to a non-hermitian interaction, $U\in\mathbb{C}$. The three-dimensional sphere is described using spherical coordinates ($\nu$ and $\eta$) via $(\varphi, \,\text{Re}(U),\,\text{Im}(U)) = (r_\varphi \cos(\nu)\sin(\eta) + \varphi_\text{a},\, r_U \sin(\nu)\sin(\eta) +U_\text{a},\, r_U\cos(\eta))$ where the endpoint (iii) is located at $\varphi_\text{a}$ and $U_\text{a}$. We highlighted the incoming EPs in black which correspond to $A_1$ and $B_1$ in Fig. \ref{fig:circles_app} and the outgoing points in red. We used a different radius compared to Fig. \ref{fig:circles_app}. $\eta = \pi/2$ corresponds to the hermitian interaction $U \in \mathbb{R}$. We illustrated the two incoming (black) and the two outgoing (red) EPs on the three-dimensional sphere in panel c) ($L=3$) and d) ($L=6$).}
\end{figure}

Finally, we find that exceptional lines survive for a non-hermitian interaction ($U\in\mathbb{C}$) after they are annihilated in the case of a hermitian interaction ($U\in\mathbb{R}$). Fig. \ref{fig:sphere_app} evaluates the minimal angle between all eigenvectors on a sphere which is parameterized using spherical coordinate ($\nu$ and $\eta$) around the endpoints (iii) in Fig. \ref{fig:circles_app}. We extent the parameter space to three dimensions by including a non-hermitian density-density interaction, $U\in\mathbb{C}$:
\begin{align}
	&(\varphi, \,\text{Re}(U),\,\text{Im}(U))\\
	=& (r_\varphi \cos(\nu)\sin(\eta) + \varphi_\text{a},\, r_U \sin(\nu)\sin(\eta) +U_\text{a},\, r_U\cos(\eta))\nonumber
\end{align}
The endpoints (iii) in Fig. \ref{fig:circles_app} are located at $\varphi_\text{a}$ and $U_\text{a}$. Panel a) and b) in Fig. \ref{fig:sphere_app} are showing the minimal angle in the $\nu$--$\eta$ plane for the system of $L=3$ and $L=6$ sites. The two incoming EPs ($A_1$ and $B_1$) are highlighted in black and are located at $\eta=\pi/2$ which corresponds to $U\in\mathbb{R}$. However, the two outgoing EPs which are marked in red have a finite imaginary part, $U\in\mathbb{C}$. Hence, they are not longer present in the case of a hermitian interaction as used in Fig. \ref{fig:circles_app}. 
Note that the incoming and outgoing points are roughly separated by $\Delta \nu=\pi$ which means they propagate on a similar trajectory but with a finite imaginary part. Also, it is worth to point out that the outgoing EPs only differ by the imaginary part of $U$, the twist angle $\varphi$ and the real part of $U$ are identical. The dotted lines in the sphere are guiding the eye and do not represent the real paths of the EPs in the three-dimensional parameter space $(\varphi, \,\text{Re}(U),\,\text{Im}(U))$.

\section{Multiple fermions}\label{app:three_fermions}

\begin{figure}
	\centering
	\includegraphics[width=\columnwidth]{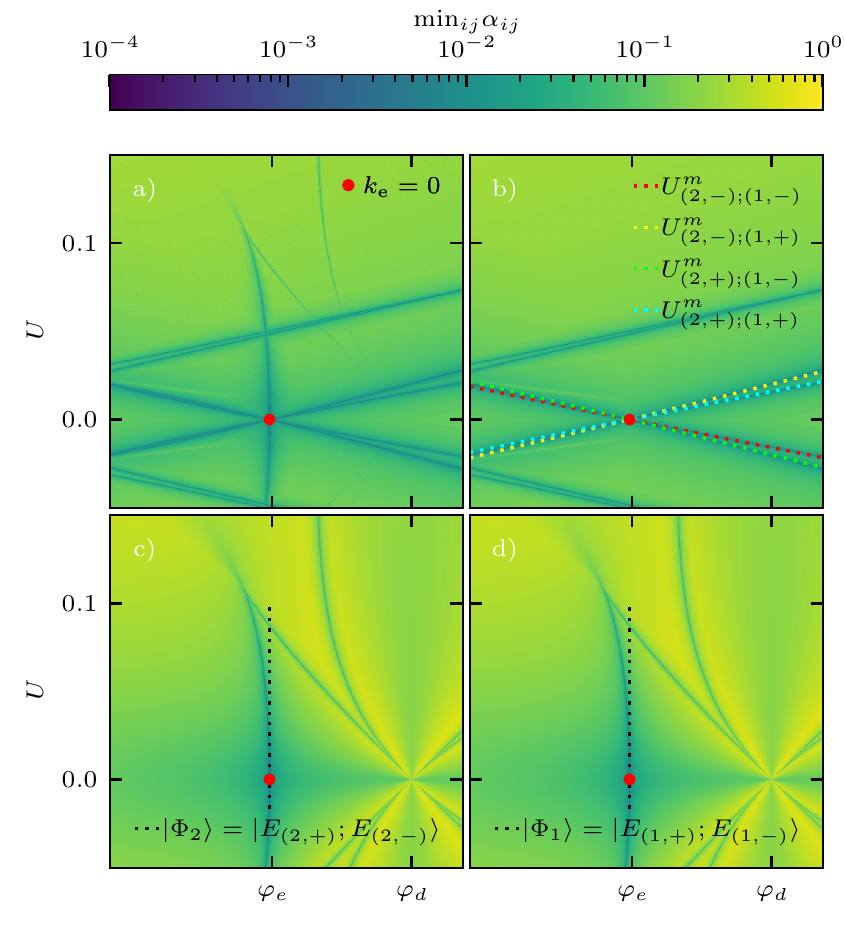}\label{fig:three_fermions}
	\caption{The figure is showing the minimal angle for a system of $L=3$ sites with $m=0.7$ with three fermions. While panel a) evaluates the full Hamiltonian, panels b), c) and d) is restricted to the total conserved momentum $k_\text{tot}=0,1,2$ respectively. EPs are inherited from the single particle spectrum at $k_{\text{e}}=0$ and $\varphi_{\text{e}}$. Panel b) includes the predicted paths of EPs using Eq. \eqref{eq:eps_stability_three_fermions_m} for a total momentum of $k_\text{tot}=0$. Panels c) and d) reveal vertical lines of EPs which originate from $\ket{\Phi_2}$ and $\ket{\Phi_1}$. As in Fig. \ref{fig:EP_prediction_app}, EPs emerge from diagonalizable degeneracies at $\varphi_\text{d}$ for $U=0$.}
\end{figure}

We briefly discuss the generalization of our perturbative expansion to three fermions.
Similar to the two-particle case, we start from the non-interacting limit and derive an effective Hamiltonian exhibiting similar physics.
We restrict the discussion to the case (i) where EPs are inherited in the same way as in the case of two interacting fermions.

The right and left eigenvector of two non-interacting particles with $k\neq q$ and $\xi_k,\xi_q = \pm$ are given in Eq. \eqref{eq:EV_k0_k1} and \eqref{eq:LEV_k0_k1}.
Let an EP be located in the single particle spectrum at $k_{\text{e}}$ and $\varphi_{\text{e}}$ such that $m_{k_{\text{e}}}=0$ or $p_{k_{\text{e}}}=0$.
Each two-particle state is combined with the two exceptional generalized eigenvectors, $\ket{a_{k_{\text{e}}}}$ and $\ket{b_{k_{\text{e}}}}$:
\begin{align}
	\ket{\Psi_\text{a}}:=&\ket{a_{k_{\text{e}}};E_{(k,\xi_k)};E_{(q,\xi_q)}}\\
	\ket{\Psi_\text{b}}:=&\ket{b_{k_{\text{e}}};E_{(k,\xi_k)};E_{(q,\xi_q)}}
\end{align}
The non-interacting effective Hamiltonian at the EP is given by: 
\begin{align}
	H_0^{\text{(i)}} =& \begin{pmatrix} E_{(k,{\xi_k})}+E_{(q,{\xi_q})} & m_{k_{\text{e}}} \\ p_{k_{\text{e}}} & E_{(k,{\xi_k})}+E_{(q,{\xi_q})} \end{pmatrix}
\end{align}
The perturbative contribution is derived from the interacting part $H_{\text{int}}$ similar to Eq. \eqref{eq:Hint0}, \eqref{eq:Hint1}, \eqref{eq:Hint2} and \eqref{eq:perturbation_general} (${c}={a},{b}$):
\

\begin{align}
	\braOket{\Psi_{\text{c}} }{ H_{\text{int}} }{\Psi_{\text{c}}} & = \frac{6E_{(k,+)}E_{(q,+)}-\xi_k\xi_q\left(m_kp_q+p_km_q\right)}{4LE_{(k,+)}E_{(q,+)}}\\
	\braOket{\Psi_{\text{a}} }{ H_{\text{int}} }{ \Psi_{\text{b}}} & = -\frac{\xi_km_kE_{(q,+)}+\xi_qm_qE_{(k,+)}}{2LE_{(k,+)}E_{(q,+)}}\\
	\braOket{\Psi_{\text{b}} }{ H_{\text{int}} }{ \Psi_{\text{a}}} & = -\frac{\xi_kp_kE_{(q,+)}+\xi_qp_qE_{(k,+)}}{2LE_{(k,+)}E_{(q,+)}}
\end{align}

As in the case for two fermions, the diagonal elements remain identical such that the Hamiltonian becomes defective if and only if $H^{\text{(i)}}_{01}=0$ or $H^{\text{(i)}}_{10}=0$ which yields
\begin{align}
	&U_{(k,\xi_k);(q,\xi_q)}^m =  \frac{2m_{k_{\text{e}}}LE_{(k,+)}E_{(q,+)}}{\xi_km_kE_{(q,+)}+\xi_qm_qE_{(k,+)}}\label{eq:eps_stability_three_fermions_m}\\
	\text{and }&U_{(k,\xi_k);(q,\xi_q)}^p =  \frac{2p_{k_{\text{e}}}LE_{(k,+)}E_{(q,+)}}{\xi_kp_kE_{(q,+)}+\xi_qp_qE_{(k,+)}}.\label{eq:eps_stability_three_fermions_p}
\end{align}

In addition to the two-particle states which are defined for two different momenta ($k\neq q$), we need to include states which are defined for a single momentum: $\ket{\Phi_{\tilde{k}}} = a_{\tilde{k}}^\dagger b_{\tilde{k}}^\dagger\ket{0}$ for $\tilde{k}\neq k_\text{e}$.
The effective matrix is particularly simple in this case as it does \textit{not} depend on $U$:
\begin{align}
	H^{\text{(i)}} =& \begin{pmatrix} 0 & m_{k_{\text{e}}} \\ p_{k_{\text{e}}} & 0 \end{pmatrix}
\end{align}
The effective Hamiltonian is defective if $m_{k_{\text{e}}}=0$ or $p_{k_{\text{e}}}=0$ which is only fulfilled for $\varphi=\varphi_{\text{e}}$. Therefore, $L-1$ additional lines of EPs run vertically for each state $\ket{\Phi_{\tilde{k}}}$ with $\tilde{k}\neq k$.

Fig. \ref{fig:three_fermions} shows the system of $L=3$ sites and $m=0.7$ for three fermions (half filling) as in Fig. \ref{fig:EP_prediction_app} for two fermions.
While panel a) shows the minimal angle for the full Hamiltonian, panel b), c) and d) show the conserved total momentum $k_\text{tot}=0,1,2$ respectively.
The exceptional momentum which hosts the EP in the single particle spectrum is $k_{\text{e}}=0$.
All two-particle states which are defined for two different momenta, $k=1$ and $q=2$, are found in panel b) with the total momentum $k_{\text{e}}+k+q=0$.
The EPs are described by Eq. \eqref{eq:eps_stability_three_fermions_m}. 
Panel c) (d)) exhibits the total momentum $k_\text{tot}=1$ ($k_\text{tot}=2$) and highlights the vertical path of the EP which is formed by $\ket{\Phi_2}$ ($\ket{\Phi_1}$).

Besides EPs which are inherited from the single particle spectrum more lines emerge from the diagonalizable degeneracy, $\varphi_\text{d}$, \cf Fig. \ref{fig:EP_prediction_app}.
In the case of two fermions, the states $\ket{E_{(0,+)};E_{(2,+)}}$ and $\ket{E_{(0,-)};E_{(2,-)}}$ are degenerated since $E_{(0,+)}=E_{(2,-)}$ at $\varphi_\text{d}$.
This generates two degenerated eigenvalues with total momentum $k_\text{tot}=1$ and $k_\text{tot}=2$.
The lower panels exhibit the same lines of EPs.
Hence, these EPs occur twice as two Jordanblocks in the full Hamiltonian, panel a).

\section{Disorder}\label{app:disorder}

\begin{figure}
	\centering
	\includegraphics[width=\columnwidth]{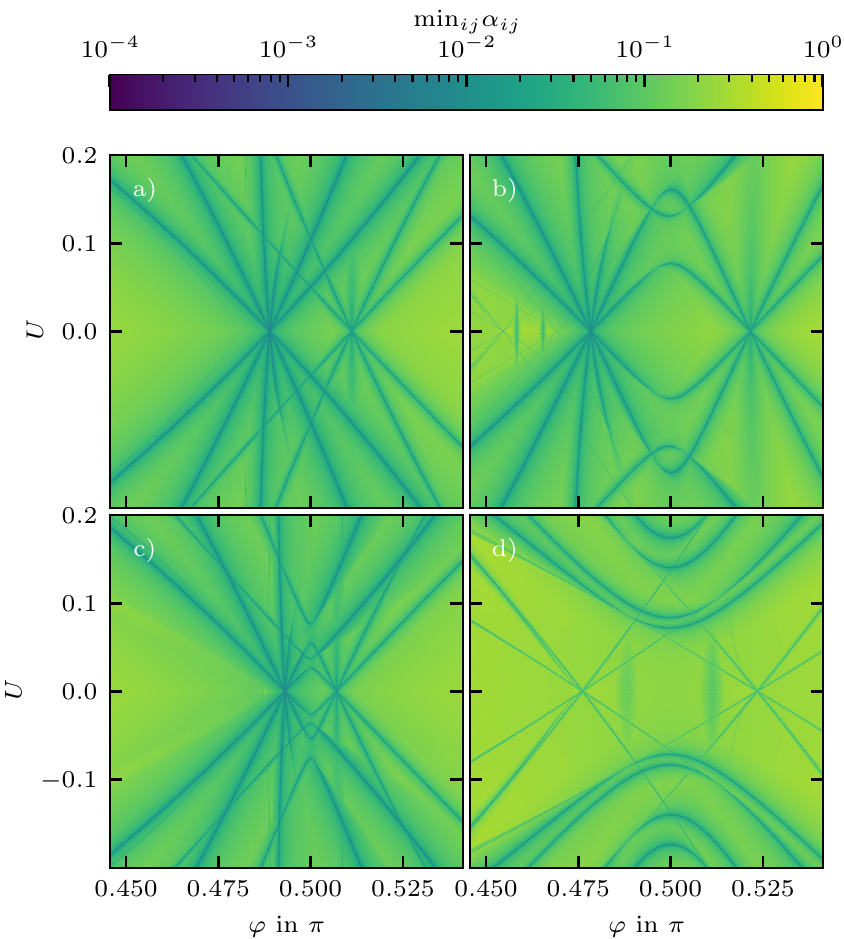}\label{fig:disorder}
	\caption{The figure is showing different disorder realization for $L=6$ sites and $m=0.7$. The disorder strength for panel a), b), c) and d) is $\delta =0.0,0.01,0.025,0.05$ respectively.}
\end{figure}

Experiments suffer from various types of disorder.
As the existence of EPs is tied to symmetries, the effect of symmetry breaking disorder is far from being obvious.
The main question is directed towards their stability and if EPs generated from interacting fermions survive in the presence of disorder.
We break translational invariance of the system by adding noise to the hopping amplitudes.
The noise is Gaussian distributed with mean value $1$ and standard deviation $\delta$.
It is multiplied to the hopping amplitudes.

Fig. \ref{fig:disorder} evaluates the response to different disorder strengths for $L=6$ sites.
EPs generated from (i) are stable but shift in momentum space.
However, EPs generated from (ii) seem to be present for small disorder strengths but show more complex patterns and vanish for larger disorder.
This is not surprising since accidental degeneracies become rare.


\clearpage
\end{document}